\documentclass[12pt,letterpaper]{article}
\pdfoutput=1 
\usepackage{jheppub}
\usepackage{amsmath}
\usepackage{amssymb}
\usepackage{graphicx}
\usepackage{array}
\usepackage{braket}
\usepackage{tikz}
\usepackage{cite}

\newcommand{\runi}{Run~1}
\newcommand{\runii}{Run~2}

\providecommand{\abs}[1]{\lvert#1\rvert}

\title{A model for the LHC diboson excess}

\author{Manuel Buen-Abad,}
\author{Andrew G. Cohen,}
\author{and Martin Schmaltz}
\affiliation{Physics Department,  Boston University\\
 Boston, MA 02215, USA}

\emailAdd{buenabad@bu.edu}
\emailAdd{cohen@bu.edu}
\emailAdd{schmaltz@bu.edu}

\abstract{The first run of the LHC showed hints of a new resonance
  with mass near $1.9$ TeV decaying into electroweak gauge boson pairs
  as well as into dijets. While \runii{} has neither confirmed nor
  ruled out such a resonance, it has yielded new constraints on models
  attempting to explain these decays.  Additionally in $W'$ models
  where this new resonance is a charged vector boson that is a weak
  isospin singlet there is the potential for conflict with the
  electroweak precision $T$ parameter. We construct variants of a $W'$
  resonance model that provide an excellent fit to both \runi{} and
  \runii{} data, as well as electroweak precision measurements.  The
  model also predicts a neutral vector boson, a $Z'$, with mass close
  to 3 TeV.  This $Z'$ is compatible with the intriguing \runii{}
  observation of a dielectron pair with invariant mass of 2.9 TeV at
  CMS.}

\begin{document} 
\maketitle
\flushbottom

\section{Introduction}
\label{sec:introduction}

Several results from
ATLAS\cite{Aad:2015owa,Aad:2014xka,Aad:2015ufa,Aad:2015yza,Aad:2014aqa}
and
CMS\cite{Khachatryan:2014hpa,Khachatryan:2014gha,Khachatryan:2016yji,
  Khachatryan:2015bma,Khachatryan:2015ywa,Khachatryan:2015sja} in
\runi{} of the LHC hint at the existence of a narrow resonance with
decays to dijet and diboson final states and a mass near 2 TeV. While
none of the individual deviations from the Standard Model (SM) have
more than 3$\sigma$ significance, the fact that several different
searches find excesses which can be explained with a single bosonic
resonance is intriguing\cite{Dias:2015mhm,Brehmer:2015dan}. There are
many interesting aspects of this data, but the feature that we will
focus on is the apparent decay of the massive resonance to electroweak
(EW) gauge boson pairs.

A particularly attractive interpretation of the \runi{} data is a new
massive charged gauge boson, a $W'$, with a mass close to 1.9 TeV
\cite{Hisano:2015gna,Franzosi:2015zra,Cheung:2015nha,Dobrescu:2015qna,
  Gao:2015irw,Thamm:2015csa,Brehmer:2015cia,Cao:2015lia,Abe:2015uaa,
  Carmona:2015xaa,Allanach:2015hba,Dobrescu:2015yba,Pelaggi:2015kna,
  Lane:2015fza,Faraggi:2015iaa,Low:2015uha,Dobrescu:2015jvn,Sajjad:2015urz,Das:2015ysz,
  Deppisch:2015cua,Dev:2015pga,Collins:2015wua}.
A combination of the \runi{} ATLAS and CMS data obtains a good fit
with a $W'$ mass close to 1.9 TeV and a $W'\to WZ$ cross section of
$5.3^{+2.3}_{-2.0}$ fb\cite{Dias:2015mhm}. \runi{} also showed
evidence for a dijet decay mode $W'\to j j$ with a cross section on
the order of 50 fb, with significant
uncertainty\cite{Brehmer:2015dan}.

Toward the end of 2015 the first results from \runii{} of the LHC at
13 TeV were announced. For most channels relating to the diboson
excess the \runii{} sensitivity was somewhat below that of \runi{},
and these new results neither confirm nor exclude the signal. A
combination of the most sensitive channels in
ATLAS\cite{ATLAS-CONF-2015-075,ATLAS-CONF-2015-071,
  ATLAS-CONF-2015-073,ATLAS-CONF-2015-068} and
CMS\cite{CMS-PAS-EXO-15-002} from \runii{} yields a 95\% exclusion
bound on the $W'\to WZ$ rate at 13 TeV for a 1.9 TeV $W'$ of 25 fb.
In addition a \runii{} ATLAS analysis places a 95\% confidence upper
limit on the dijet rate of about $150$ fb\cite{ATLAS:2015nsi}. The
parton luminosities appropriate for $W'$ production are approximately
6 times larger at 13 TeV compared to 8 TeV, and thus these \runii{}
limits correspond to 95\% confidence \runi{} limits of
$\sigma_{WZ} < 4 \text{ fb and } \sigma_{jj} < 25$ fb.

A charged $W'$ gauge boson of this sort must arise from a non-abelian
group including $SU(2)$ and therefore comes with a neutral partner, a
$Z'$.  We imagine an effective theory below some scale
$f \gg 246 \text{ GeV}$ in which the unbroken SM gauge group
$SU(2)_{SM}\times U(1)_{Y}$ is supplemented by the new massive gauge
bosons, one or more
Higgs doublets, and where all
operators of dimension greater than four are suppressed by the high
scale. To produce the diboson signal the $W'$ should decay to pairs of
EW gauge bosons. This allows two possibilities for the quantum numbers
of $V'={W', Z'}$ under the SM gauge group $SU(2)_{SM}\times U(1)_{Y}$:
\begin{itemize}
\item $V'$ is a triplet under $SU(2)_{SM}$ and has zero hypercharge.
  We refer to the resulting massive vector bosons as ``left-handed''
  and this model as the ``left-handed'' model.
\item $V'$ is a singlet under $SU(2)_{SM}$ and the hypercharges are
  $\pm{1}, 0$.  We refer to the resulting massive vector bosons as
  ``right-handed'' and this model as the ``right-handed'' model. This
  case is the focus of this work.
\end{itemize}

There are no renormalizable, gauge-invariant operators in this
effective theory that couple the $V'$ to pairs of SM gauge bosons and
produce the diboson signal.  We can obtain the desired decay of the
massive resonances through higher dimension operators, but these are
generically too small.  Alternatively the massive resonance may couple
to the longitudinal components of the $W$ and $Z$ after EW symmetry
breaking. That is, the scalar fields that acquire EW vevs (and contain
the longitudinal components of the $W$ and the $Z$) can couple at
dimension four to the massive resonances. In our effective theory the
only relevant operators take the form of the massive gauge bosons
times currents constructed from the Higgs fields.  For the left-handed
case these currents must be $SU(2)_{SM}$ triplets and $U(1)_{Y}$
singlets, while for the right-handed case they must by $SU(2)_{SM}$
singlets with hypercharge $\pm 1,0$. We denote them generically as
\begin{equation}
  \label{eq:1}
  g_{V'} V'_\mu 
  \Phi i D^\mu \Phi
\end{equation}
where $g_{V'}$ is a coupling constant, $D$ is the covariant derivative
including the electroweak gauge fields, $\Phi$ is a Higgs field (or
its conjugate) and we have suppressed explicit indices.  The form of
these operators is one of the reasons that a heavy vector resonance
with a diboson decay mode is of such interest: this decay is a direct
measurement of EW symmetry breaking and probes the details of the
Higgs vevs.

In addition to providing the diboson decay, the operator \eqref{eq:1}
includes mass mixing of the heavy resonances with the $W$ and the
$Z$. This mixing may shift the mass of the $W$ relative to the
$Z$. While this shift is small, the extraordinarily precise measured
values of these masses significantly constrain such an effect:
electroweak precision measurements preclude a large correction to the
$T$ parameter. There is a straightforward way to help protect against
such a correction: incorporate a custodial $SU(2)$ symmetry. This is
automatic in the left-handed model where the heavy resonances are a
triplet under $SU(2)_{SM}$. The right-handed model has no such
protection: the $W'$ and $Z'$ are not members of an $SU(2)_{SM}$
triplet and their mixing with the $W$ and $Z$ violates custodial
$SU(2)$.  Therefore the operators responsible for the heavy vector
decay into dibosons may also generate a non-zero value of the $T$
parameter.  This is the main topic of our paper: exploration of the
tension between the constraints on the $T$ parameter and the diboson
branching fraction for models with a right-handed $W'$. We will find
that relaxing this tension prefers a $Z'$ mass right around 3 TeV.

Upon substituting EW symmetry breaking vevs and allowing for
independent couplings of the heavy $W'$ and $Z'$ resonances the
operators of \eqref{eq:1} correspond to the mixing terms
\begin{equation}
  \label{eq:2}
  \kappa_{W} M_{W}^{2}  {W'}^{-}W^{+} +\text{h.c.} + \kappa_{Z}
  M_{Z}^{2} \cos\theta_{W} Z'Z 
\end{equation}
where we have parameterized the couplings relative to the electroweak
gauge boson masses and the electroweak mixing angle
($\cos\theta_{W}\equiv M_{W}/M_{Z}$) for convenience. 

The coupling $\kappa_{W}$ determines the rate for $W'$ decay to $W Z$
\begin{equation}
  \label{eq:3}
  \Gamma(W'\to W Z) = \abs{\kappa_{W}}^{2} \frac{g^2}{192 \pi} M_{W'} 
\end{equation}
where $g$ is the $SU(2)_{SM}$ gauge coupling.  The corresponding rate
measured at the LHC is the product of the $W'$ production rate times
the branching fraction of the $W'$ into $W Z$.  Significant production
of the $W'$ requires a coupling $g_{ud}$ to the first family of
quarks. In a straightforward implementation of an $SU(2)_{M}$ gauge
theory in which the quarks are doublets under
$SU(2)_{M}$\cite{Dobrescu:2015jvn,Coloma:2015una,Dobrescu:2015yba,Dobrescu:2015qna,Brehmer:2015cia},
$g_{ud}$ is simply the gauge coupling $g_{M}$. The $W'$ then couples
universally to all three families of quarks, with a decay rate
\begin{equation}
  \label{eq:4}
  \Gamma(W'\to q\bar q)=3\frac{g_{M}^{2}}{16 \pi} M_{W'} .  
\end{equation}
However the rate for $W' \to W Z$ is determined by the same gauge
coupling times a factor for the fraction of the longitudinal $W$ and
$Z$ bosons contained in the scalar field $\Phi$. This fraction is
necessarily less than one, and therefore the $W Z$ decay rate is
bounded by $\Gamma(W'\to W Z) \le M_{W'} g_{M}^{2}/(192\pi)$. This
leads to a lower bound on the dijet rate relative to the $W Z$ event
rate:
\begin{equation}
  \label{eq:5}
  \sigma_{jj} \ge  36 \, \sigma_{WZ}
\end{equation}
A \runi{} $W Z$ signal of a few femtobarns thus requires a \runi{}
dijet rate in excess of a hundred femtobarns. Such a large dijet rate
is fully excluded by the \runii{} data. For this reason the models we
construct will incorporate fermion mixing, allowing the $W'$ coupling
to first family quarks $g_{ud}$ to differ from the gauge coupling
$g_{M}$.  Mixing of fermions inevitably involves issues of flavor, and
without fine tuning or additional flavor symmetries we run the risk of
significant flavor changing neutral currents. We therefore include
flavor symmetry to afford some protection against these dangerous
effects.  We will consider two examples: one in which the coupling to
all three families is universal; and another with universal couplings
to first and second families but no coupling of the $W'$ to the third.

It is convenient to write the $W Z$ branching fraction in terms of the
branching fraction of the $W'$ to quarks. Defining
$B_{jj}\equiv \text{B}(W'\to q\bar q)$ this is
$\text{B}(W'\to W Z) = B_{jj} \cdot \Gamma(W'\to W Z)/\Gamma(W'\to
q\bar q)$.\footnote{An additional diboson signal stems from the $W'$
  decay to Higgs particles, $W'\to W h$. In models with a single Higgs
  doublet the rate for this mode is equal to that of $W'\to W Z$. With
  multiple Higgs doublets the rates may differ. However since the
  observed Higgs particle has couplings consistent with the full vev
  of $246$ GeV this suggests that the observed Higgs couples to $W'$
  like the full vev as well. In this case the $W'\to W h$ rate is
  again the same as the $W'\to W Z$ rate.}  With $N_{f}$ the number of
families that the $W'$ couples to (either 2 or 3) the decay rate to
quarks is
\begin{equation}
  \label{eq:6}
  \Gamma(W'\to q\bar q)=N_{f}\frac{g_{ud}^{2}}{16 \pi} M_{W'} .
\end{equation}

The rate for $W'$ production may be computed by integrating the
production cross section over parton distribution functions
\begin{equation}
  \label{eq:7}
  \sigma(p p\to W') =
  \frac{\pi}{6} \frac{g_{ud}^2}s \int_{M_{W'}^2/s}^1
  \frac{dx}{x}\left[f_u(x) f_{\bar d}(\frac{M_{W'}^2}{x s})+f_d(x)
    f_{\bar u}(\frac{M_{W'}^2}{x s})
  \right] \simeq  g_{ud}^{2} \, 0.8
  \text{ pb} \ . 
\end{equation}
Here $\sqrt{s}=8$ TeV is the collider center of mass energy, the
$f_i(x)$ are the parton distribution functions, and we have summed
over both first and second family quarks. For our numerical results we
use MSTW parton distribution functions \cite{Martin:2009iq} with NLO
K-factors taken from \cite{Cao:2012ng,
    Carena:2004xs, Hamberg:1990np}.  The diboson cross section
from $W'$ production $\sigma_{WZ}$ is then the product
\begin{multline}
  \label{eq:8}
  \sigma_{WZ}=\sigma(pp\to W') \text{ B}(W'\to W Z) = \\
  \sigma(pp\to W') B_{jj} \frac{\Gamma(W'\to W
    Z)}{\Gamma(W'\to q\bar q)} = \\
  \abs{\kappa_{W}}^2 \frac{B_{jj}}{N_{f}}\, \frac{g^2}{12} \, 0.8 
  \text{ pb} =  \abs{\kappa_{W}}^2 \frac{B_{jj}}{N_{f}} \, 28 \text{ fb}.
\end{multline}
Note that the dependence on the fermion coupling to the $W'$ has been
subsumed in the branching fraction to quarks. We then have a
prediction for $\sigma_{W Z}$ with $\kappa_{W}$ and $B_{jj}/N_{f}$ as
the only free parameters. As a rough benchmark, a signal of
$4 \text{ fb}$ with $N_{f}=2$ corresponds to
$B_{jj} \abs{\kappa_{W}}^{2} \simeq .29$.  Since $B_{jj} < 1$ this
means that $\kappa_{W} \gtrsim .5$ to obtain this cross section.

We may develop some intuition for the precision electroweak
constraints that apply to~\eqref{eq:2} by noting that the most
precisely measured electroweak parameters are the Fermi constant
$G_{F}$, the fine structure constant at the $Z$ mass $\alpha(M_{Z})$,
the mass of the $Z$, $M_{Z}$, and the mass of the $W$, $M_{W}$. In the
SM any three of these observables may be used to fix the parameters in
the gauge sector of the theory ($g, g', v$) and then one prediction
for the remaining parameter may be obtained.\footnote{This prediction
  is only weakly dependent on the other parameters of the model, such
  as $\alpha_{s}$, the Higgs mass and quartic coupling, and the quark
  Yukawa couplings.} The same procedure may be applied including the
operators of~\eqref{eq:2} where now the prediction depends on the
parameters $\kappa_{W,Z}$. It is convenient to phrase this prediction
as $M_{W}^{2}/M_{W0}^{2} - M_{Z}^{2}/M_{Z0}^{2}$ where the subscript
$0$ indicates the SM value. The SM prediction for this parameter is
clearly zero, whereas in the resonance model we need only compute
shifts in masses from SM values:
$\delta M_{W}^{2}/M_{W0}^{2} - \delta M_{Z}^{2}/M_{Z0}^{2}$. In the
absence of couplings of the heavy resonances other than those
in~\eqref{eq:2} this is just the conventionally defined $T$
parameter\cite{Agashe:2014kda} given by (to leading order in inverse
powers of the heavy masses)
\begin{equation}
  \label{eq:9}
  \alpha(M_{Z}) T   = -\abs{\kappa_{W}}^{2} \frac{M_{W}^{2}}{M_{W'}^{2}} +
  \kappa_{Z}^{2} \cos^{2}\theta_{W}\frac{M_{Z}^{2}}{M_{Z'}^{2}}  =
  \frac{M_{W}^{2}}{M_{W'}^{2}} 
  \left\{
    -\abs{\kappa_{W}}^{2} + \kappa_{Z}^{2} \frac{M_{W'}^{2}}{M_{Z'}^{2}}
  \right\}
\end{equation}
There are no tree level contributions to the other prominent
electroweak precision parameter $S$. Precision measurements constrain
$\alpha(M_{Z}) T$ to be less than $10^{-3}$ which implies
$\kappa_{W,Z}$ no larger than of order one, or cancellations between
the $W'$ and $Z'$ contributions. With $S=0$ the precision fit has a
preference for positive values of $T$,
$\alpha T = (4 \pm 2.4) \times 10^{-4}$, suggesting that the $Z'$
contribution should be larger than that of the $W'$.

The inherent tension between the diboson signal and the $T$ parameter
is already evident in these general expressions. A large diboson rate
requires a large value for $\kappa_{W}$ ($B_{jj}$ can only suppress
the rate), but this pushes the $T$ parameter in the wrong
direction. This can be compensated by a contribution from the $Z'$
through $\kappa_{Z}$, but only if the $Z'$ is not too heavy.

To refine this constraint on the $Z'$ mass we need to make some
choices.  We may construct a right-handed $W'$ model starting with the
gauge group $SU(2)_{L}\times SU(2)_{M}\times U(1)_{X}$ and breaking
$SU(2)_{M}\times U(1)_{X} \to U(1)_{Y}$ at the scale $f$.  The
resulting massive gauge bosons are the right-handed
$W' \text{ and } Z'$ and below this scale we have the desired
effective theory. Including a $(1,R)_{(R-1)/2}$ field $H_{XM}$ (for
some non-trivial representation of dimension $R$) that acquires a
large vev $f/\sqrt{2}$ accomplishes the desired breaking. The
hypercharge gauge coupling is
$g' = g_{M} g_{X}/\sqrt{g_{M}^{2}+g_{X}^{2}} \equiv g_{M}
\sin\theta_{M}$. The smallest such representation is $R=2$, the
doublet. However as we will see this leads to either a poor precision
fit or fine tuning. Therefore we prefer an $SU(2)_{M}$ triplet,
$R=3$. In this case the $Z'$ mass is
$M_{Z'} = \sqrt{2} M_{W'}/\cos\theta_{M}$.

\begin{figure}[ht]
  \centering
  \begin{tikzpicture}[scale=0.9, transform shape]
    \node (L) [draw, circle] at (0, 0) {$SU(2)_{L}$};
    \node (X)  [draw, circle, fill=gray!25]  at +(0: 5) {$U(1)_{X}$}; 
    \node (M) [draw, circle] at +(60: 5) {$SU(2)_{M}$};
     \draw (L) -- (M) node [pos=.5,sloped,above] {$H_{M}$};
    \draw (X) -- (M) node [pos=.5,sloped,above] {$H_{XM}$}; 
    \draw (L) -- (X) node [pos=.5,sloped,above] {$H_{X}$};
  \end{tikzpicture}
  \caption{Theory space diagram representing the bosonic field content
    for the models described in the text.}
  \label{fig:1}
\end{figure}
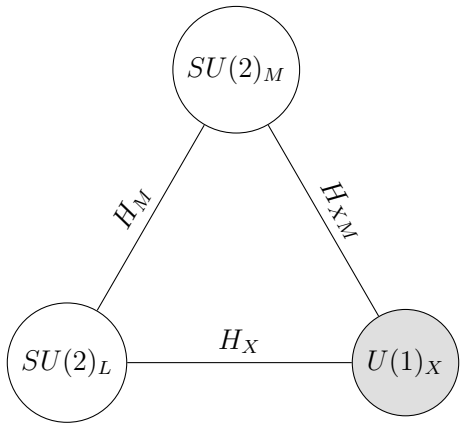

We must also include scalar field representations that contain Higgs
doublets following this breaking. These are representations of the
form $(2,R)_{X}$. The smallest such representations, each containing
four real fields, are a complex doublet $(2,1)_{1/2}$ field $H_{X}$
and a real ``bi-doublet'' $(2,2)_{0}$ field $H_{M}$.\footnote{Since
  $SU(2)\times SU(2) \sim SO(4)$ the real bidoublet may be
  equivalently thought of as the vector of $SO(4)$. We may represent
  this field in a variety of ways: as a complex two component vector
  $\bigl( \begin{smallmatrix}  \phi^{+} \\
    \phi^{0} \end{smallmatrix}\bigr)$ (as we choose here); as a
  $2\times 2$ matrix $
  \bigl( \begin{smallmatrix} {\phi^{0}}^{*} & \phi^{+} \\
    -{\phi^{+}}^{*} & \phi^{0}
  \end{smallmatrix} \bigr)$; or as a 4-component column vector formed
  from the real and imaginary parts of
  $\phi^{0}\text{ and } \phi^{+}$.}  The model including these
representations is nicely summarized by the theory space diagram of
Figure \ref{fig:1}.

The mixing operators in~\eqref{eq:1} then take the form
\begin{multline}
  \label{eq:12}
  \frac{g_{M}}{2}\sqrt{2} {W'}^{-}_{\mu}\, H^{T}_{M} i\sigma^{2} i
  D^{\mu} H_{M} + \text{h.c.}  \\
  + \frac{g_{M}}{2}\cos\theta_{M} Z'_{\mu} H_{M}^{\dagger} iD^{\mu}
  H_{M} - \frac{g_{X}}{2}\sin\theta_{M}
  Z'_{\mu}   H_{X}^{\dagger} iD^{\mu} H_{X}
\end{multline}
Our models will include two of the bidoublets $H_{M}$ and, for the
moment, we will ignore any vevs for the $H_{X}$ fields.  Each of these
bidoublets is an SM Higgs field and we therefore have a multi-Higgs
doublet model.

The sum $\sum \abs{v_{M}}^{2}+\abs{v_{X}}^{2} \equiv v^{2}$ is
constrained to the electroweak value $v^{2}=(246 \text{ GeV})^{2}$.
Note that the phases in the vevs of these bidoublets are not
necessarily aligned and
$t^{2} \equiv \abs{\sum v_{M}^{2}} / v^{2} \le 1$.  The $W'$ mixing
in~\eqref{eq:12} is sensitive to these phases and thus $t$ appears in
$\kappa_{W}$. Both $\kappa_{W}$ and $\kappa_{Z}$ are readily computed
(ignoring $v_{X}$)
\begin{equation}\label{eq:14}
  \begin{split}
    \kappa_{W} &=  \frac{g_{M}}{g}t^{2} = \frac{\tan\theta_{W}}{\sin\theta_{M}} t^{2}  \\
    \kappa_{Z} &=  \frac{g_{M}}{g}\cos\theta_M=
    \frac{\tan\theta_{W}}{\sin\theta_{M}} \cos\theta_M 
  \end{split}
\end{equation}
so that~\eqref{eq:9} gives
\begin{equation}
  \label{eq:15}
  \alpha(M_{Z}) T =
  \frac{\tan^{2}\theta_{W}}{\sin^{2}\theta_{M}}\frac{M_{W}^{2}}{M_{W'}^{2}}
  \left[
    \frac{\cos^4\theta_M}{2} - t^{4}
\right]
\end{equation}

When the phases of the Higgs fields are all aligned (or in the case of
a single bidoublet field in which case the phase is necessarily
aligned) $t=1$ and~\eqref{eq:15} gives
\begin{equation}
  \label{eq:16}
  \alpha(M_{Z}) T =
  -\tan^{2}\theta_{W}\frac{M_{W}^{2}}{M_{W'}^{2}} 
  \left(
    \frac{2-\cos^{4}\theta_{M}}{2\sin^{2}\theta_{M}}
  \right)  \lesssim -5.3 \times 10^{-4} 
\end{equation}
where we have set $M_{W'} = 1.9 \text{ TeV}$.  Although the $T$
parameter in~\eqref{eq:16} is not much larger than the experimental
uncertainty, it is unfortunately negative, and more than 3 sigma away
from the experimental value. The negative definiteness of the result
reflects the dominance of the $W'$ contribution over that of the $Z'$
for all values of the $Z'$ mass. Evading the constraint in
\eqref{eq:16} is necessary for a good precision fit, and \eqref{eq:15}
demonstrates that this requires non-aligned vevs so that $t^{2} < 1$.

With these ingredients in place, \eqref{eq:8} and \eqref{eq:15} relate
$\alpha T$, $\sigma_{W Z}$, $B_{jj}/N_{f}$ and the $Z'$ mass:
\begin{equation}
  \label{eq:17}
  2\frac{M_{W'}^{4}}{M_{Z'}^{4}}\frac{1}{1-2M_{W'}^{2}/M_{Z'}^{2}}
  = \alpha(M_{Z}) T
  \frac{1}{\tan^{2}\theta_{W}} \frac{M_{W'}^{2}}{M_{W}^{2}} +
  \frac{N_{f}}{B_{jj}} \frac{\sigma_{W Z}}{28\text{ fb}}
\end{equation}

We may trade $B_{jj}$ for the dijet resonance cross section at $1.9$
TeV. Assuming all decays aside from dijets and dibosons are small
\begin{equation}
  \label{eq:18}
  B_{jj} =  \frac{\sigma_{jj}}{\sigma_{jj}+2\sigma_{WZ}}
\end{equation}
where we have used $\sigma_{WZ}+\sigma_{Wh} \simeq 2 \sigma_{WZ}$.

In fact $M_{Z'}$ as determined by \eqref{eq:17} and \eqref{eq:18} is
rather insensitive to the values of the dijet cross section preferred
by the data. For $\sigma_{jj}$ much larger than
$\sigma_{WZ}\sim 4 \text{ fb}$ the dijet branching fraction
$B_{jj}\simeq 1$ and any dependence on $\sigma_{jj}$
disappears. Smaller values of $\sigma_{jj}$ suppress $B_{jj}$ and
require larger values of $\kappa_{W}$, in turn requiring a smaller
$Z'$ mass to fit the $T$ parameter. Note that for very small values of
the dijet cross section, $\sigma_{jj} \ll 1 \text{ fb}$, the last term in
\eqref{eq:17} grows large, and avoiding unacceptably large corrections
to the $T$ parameter requires fine tuning of the $Z'$ mass such that
the left hand side of this equation compensates. To avoid this tuning
we will prefer parameters which yield a dijet cross section greater
than a few femtobarns.

Fixing
$M_{W'} = 1.9 \text{ TeV}, \sigma_{WZ}=4 \text{ fb}, \sigma_{jj} > 1
\text{ fb}$, and allowing $\alpha T$ to vary over its 1~$\sigma$ range
we find a range for the $Z'$ mass of
\begin{equation}
  \label{eq:19}
  2.8 \text{ TeV} < M_{Z'} < 3.2 \text{ TeV}
\end{equation}
We will refine this analysis by performing a full precision fit in the
next section, but it is clear that the dominant driver of a light $Z'$
is the $T$ parameter, and a $Z'$ mass close to 3 TeV is necessary for
a good fit.

How does this result depend on the model choices made? As already
remarked the dijet cross section does not make much difference, and
therefore choosing non-universal couplings of the $W'$ to the first
two families (which only enter through the dijet cross section) or
including a significant coupling to the third generation makes only a
small difference.  We might also contemplate other decay modes of the
$W'$, such as to leptons (with a light right-handed neutrino), Higgs
scalars, or new fermions. Such decays would lower $B_{jj}$ which in
turn requires a lighter $Z'$. The $Z'$ mass is bounded from below,
$M_{Z'} \ge \sqrt{2} M_{W'} \simeq 2.69 \text{ TeV}$, and as $B_{jj}$
gets very small the $Z'$ mass approaches this value.

More significantly, we might have chosen the breaking of
$SU(2)_{M}\times U(1)_{X}$ through an $R$ dimensional representation
other than a triplet. Choosing a doublet would give a lighter $Z'$,
but a somewhat worse precision fit. In addition such a light $Z'$
comes with restrictive direct experimental bounds, predominantly from
the $Z'$ decay to leptons. Evading these bounds requires some fine
tuning. For these reasons we prefer the triplet. Higher dimensional
representations are also possible, and yield good precision fits with
larger $Z'$ masses, although with larger coupling $g_{M}$.

Finally we may consider including a significant vev for the alternate
Higgs representation, $H_{X}$. The total vev squared of all Higgs
doublets is fixed at $246$ GeV, and including a larger vev for $H_{X}$
necessitates decreasing the vev for the bidoublets.  Since the diboson
decay of the $W'$ comes only from the bidoublet vevs, a large diboson
rate precludes a large value for the $H_{X}$ vev. Consequently the
presence of this vev has only a small effect on the $Z'$ mass. As we
will see in our full model fits the trend is to push the $Z'$ mass to
the low end of the range \eqref{eq:19}.

\section{Symmetry breaking}
\label{sec:higgses}

In this Section we summarize the properties of the scalar fields with
vacuum expectation values that result in spontaneous breaking of the
gauge invariances. For the right-handed $W'$ physics that we are
considering there are two categories of scalars: those that break
$SU(2)_{M}\times U(1)_{X}\to U(1)_{Y}$ at the high scale $f$, allowing
the $W'$ and $Z'$ to acquire large masses, and those that implement
the SM breaking $SU(2)_{SM}\times U(1)_{Y}\to U(1)_{Q}$ at the scale
$v$.

The breaking at the scale $f$ is accomplished by a (set of) complex
scalar(s) $H_{XM}$ transforming as $(1,R)_{(R-1)/2}$ under
$(SU(2)_{L}, SU(2)_M)_{U(1)_X}$. Here $R$ is the dimension of the
isospin representation of $SU(2)_M$ and the $U(1)_X$ charge is
adjusted to preserve the conventionally chosen hypercharge generator
$Y=T_M^3+X$. For any $R>1$ the unbroken gauge group is the Standard
Model. A conventional normalization for the vacuum expectation values
 leads to masses for the $W'$ and
$Z'$
\begin{equation}\label{eq:20}
  \begin{split}
    M^2_{W'} &=\frac{g_M^2}{4} \sum_{R>1} f_R^2 = \frac{g'^2}{4\sin^2\theta_M}
    \sum_{R>1} f_R^2  \\ 
    M^2_{Z'} &=\frac{g_M^2+g_X^2}{4} \sum_{R>1} (R-1) f_R^2
    =\frac{g'^2}{4\sin^2\theta_M\cos^2\theta_M} \sum_{R>1} (R-1) f_R^2 \ .  
  \end{split}
\end{equation}
We limit ourselves to $SU(2)_M$ doublet and triplet
representations. One of our results is that the precision fit prefers
triplet breaking so that $M_{Z'} = \sqrt{2}
M_{W'}/\cos\theta_M$. However a doublet is needed to adequately
account for quark masses and Yukawa couplings. Consequently we will
include both representations in our models, with a small doublet vev
that modifies this mass relation by a few percent.

The subsequent breaking of EW symmetry at the scale $v$ must come
(predominantly) from fields that transform as doublets under
$SU(2)_{SM}$ with hypercharge $\pm 1/2$. As discussed earlier there
are two small representations of the full gauge theory that contain
Higgs doublets following the breaking at the scale $f$, and we include
them both: fields $H_X$ transforming as $(2,1)_{\frac12}$ and fields
$H_M$ transforming as $(2,2)_0$. Both fields transform as ordinary
Higgs doublets under the SM gauge group and preserve the usual leading
order mass relation $M_Z=M_W/\cos\theta_W$. However the two types of
representations have different couplings to the $W'$ and $Z'$. We have
already discussed the consequences of this for $W W'$ and $Z Z'$
mixing and the associated effects on the $T$ parameter.  Integrating
out the heavy gauge bosons yields additional dimension six operators
that are sensitive to the choice of representation and contribute to
the precision electroweak fit. The relevant couplings of both types of
Higgs doublets to the $W'$ and $Z'$ are given in \eqref{eq:12}.

\section{Fermion masses and mixings}
\label{sec:yukawas}

Obtaining satisfactory predictions for the SM fermion masses and CKM
matrix without also generating excessive flavor changing neutral
current (FCNC) couplings and meson anti-meson mixings is notoriously
difficult in models with right-handed $SU(2)$ gauge bosons. CP
violation in Kaon mixing leads to especially strong constraints on the
couplings of the new states to first and second family quarks. Most
dangerous are FCNC couplings of the $Z'$, the Higgs, and box diagrams
with exchange of one $W$ and one $W'$ boson involving first and second
family quarks.

These dangerous flavor changing effects can be avoided altogether if
the right-handed SM quarks are singlets under $SU(2)_{M}$, in which
case fermion masses and mixings may be introduced through Yukawa
couplings exactly as in the SM. The gauge couplings of the fermions
preserve a full $U(3)^5$ flavor symmetry and flavor violation enters
only through these Yukawa couplings, also exactly as in the
SM. However explaining the diboson anomaly requires a significant $W'$
coupling to first family quarks in order to adequately produce the
$W'$. Thus the up and down quark must be at least partially contained
in a doublet of $SU(2)_M$. First family quark couplings to $W'$ and
$Z'$ bosons are then proportional to mixing angles of quark singlets
with these doublets.  In order to minimize FCNCs relevant to Kaon
physics we assume that these mixing angles respect (at least)
$SU(2)$-flavor symmetries acting on the first and second family
quarks.

For our precision fits the details of the fermion mass and Yukawa
terms in the Lagrangian are largely irrelevant: we only need the
couplings of fermions to the $W'$ and the $Z'$. These are determined
by the fermion charges and the fraction of each SM quark that is
$SU(2)_{M}$ doublet.  Introducing mixing angles for each SM fermion
$s_{f} \equiv \sin\theta_{f}$, (where $s_{f} = 0$ corresponds to pure
$SU(2)_{M}$ singlet fermions), these couplings take a simple generic
form. The coupling of the $W'$ to the $SU(2)_{SM}$ singlet up and down
quarks is
\begin{equation}
  \label{eq:21}
  s_u s_d\, g_M = s_{u} s_d \frac{g'}{\sin\theta_M} \ ,  
\end{equation}
and equivalently for $(c,s)$ and $(t,b)$. We assume that $SU(2)_{SM}$
singlet neutrinos are heavy and there are no relevant couplings of the
$W'$ to the SM leptons. For the $Z'$ coupling to a SM fermion field $f$ we have
\begin{equation}
  \label{eq:22}
  g_f=\frac{g'}{\sin\theta_M\cos\theta_M} (s_f^2 T^{3}_{M} -
  \sin^2\theta_M Y)
\end{equation}
where $s_f$ is the fermion mixing angle, $T^{3}_{M}$ is the
$SU(2)_{M}$ isospin of the fermion field $f$ and $Y$ is the usual SM
hypercharge.

In the following we describe two example models with different flavor
symmetries. Obtaining a large top quark mass in extensions of the SM
is often a challenge, and our first model will treat the third family
differently from the first two. For the first two families we
implement an approximate $SU(2)$ flavor symmetry and obtain the SM
quark masses through couplings to the $H_{X}$ field which has a small
vev, while for the top quark we couple to the bidoublet fields with
their larger vevs.

Our second model realizes the top mass through coupling only to the
field $H_{X}$ and we impose an approximate $SU(3)$ flavor symmetry on
all three families of quarks. As we will see, this model also provides
an excellent fit to the data, albeit at the expense of larger coupling
constants and some modest tuning of parameters.

While we do not give a specific implementation for the lepton sectors,
it is straightforward to extend the kind of structures we present for
the quarks to leptons. In both models we will assume a separate
approximate $SU(3)$ flavor symmetry on the leptons and then, for the
purposes of this paper, the only parameter that enters the lepton
phenomenology is a universal lepton mixing angle.

\subsection{$SU(2)$ flavor model}

\begin{table}[htb]
  \centering
  \begin{tabular}{>{$}c<{$}|>{$}c<{$}|>{$}c<{$}|>{$}c<{$}|>{$}r<{$}}
    &  SU(3)_{c} & SU(2)_{L} & SU(2)_{M} & U(1)_{X} \\
    \hline{}
    q & 3 &2 & 1& \frac16{}\\
    U^c & \bar{3} &1 & 1& -\frac23{}\\
    D^c & \bar{3} &1 & 1& \frac13{}\\
    Q & 3 & 1 & 2 & \frac16{}\\
    Q^c & \bar{3} & 1 & 2 & -\frac16{}\\
  \end{tabular}
  \hskip 1cm
  \begin{tikzpicture}[scale=0.65, baseline=1.5cm, transform shape]
    \node (L) [draw, circle] at (0, 0) {$SU(2)_{L}$};
    \node (X)  [draw, circle, fill=gray!25, label=+5:$U^{c}$, label=-5:$D^{c}$]  
    at +(0: 5) {$U(1)_{X}$}; 
    \node (M) [draw, circle] at +(60: 5) {$SU(2)_{M}$};
    \draw (L) -- (M) node [pos=.5,sloped,above] {$H_{M}$};
    \draw (X) -- (M) node [pos=.5,sloped,above] {$H_{XM}$} node
    [pos=.5,sloped,below] {$Q, Q^{c}$}; 
    \draw (L) -- (X) node [pos=.5,sloped,above] {$H_{X}$} node
    [pos=.5,sloped,below] {$q$}; 
  \end{tikzpicture}
  \caption{Fields and charges for the $SU(2)$ flavor model. All fields
    are left-handed.}
  \label{tab:1}
\end{table}
In our first model we take both up- and down-type anti-quarks to be
admixtures of $SU(2)_M$ singlets and doublets, and assume that this
mixing respects $SU(2)$ flavor symmetries acting on the first and
second family anti-quark fields.  We treat the third family
separately, making the top quark pure $SU(2)_M$ doublet and the bottom
quark pure singlet, easily accommodating a large top quark mass.

The quarks of the first two families along with their vector-like
partners are described by the fields given in Table
\ref{tab:1}. Yukawa couplings and masses for the heavy fermions stem
from the Lagrangian
\begin{equation}
  \label{eq:23}
  \mathcal{L} \supset
  y^u q H_X U^c + y^{d} q \tilde H_X D^c + Y^{u} Q H_{XM} U^c +
  Y^{d} Q \tilde H_{XM} D^c + m Q Q^c \ .  
\end{equation}
Here $H_{XM}$ is an $SU(2)_{M}$ doublet whose vev is at the TeV
scale. The mass scale $m$ is also assumed to be at the TeV scale.  At
this scale a linear combination of the fields $(U^c,D^c)$ and $Q^c$
obtain a large Dirac mass with the field $Q$ from the last three terms
in \eqref{eq:23}.  Assuming that these terms respect the $SU(2)$
flavor symmetry the mixing angle which parameterizes this linear
combination is universal for the first two families.  The orthogonal
linear combinations of $(U^c,D^c)$ and $Q^c$ correspond to the SM
anti-quarks.  They obtain their Yukawa couplings to $q$ and the EW
breaking Higgs doublet $H_X$ from the first two terms.  Since the
masses of the first two families are very small, the vev of $H_X$ can
be a subdominant source of EW breaking $v_X \ll v$. This allows the
majority of the breaking to come from the vevs of the bidoublets $H_M$
which determine the vector boson mixing parameters $\kappa_W$ and
$\kappa_Z$. We may then use these large bidoublet vevs to obtain the
top quark mass.

The third family quarks and their masses arise from the additional
Lagrangian
\begin{equation}
  \label{eq:24}
  \mathcal{L} \supset y^t q_{3} H_{M} Q_{3}^c + y^{b} q_{3} \tilde{H}_X D_{3}^c + y^{Q}
  D_{3} \tilde{H}_{XM} Q_{3}^c  \ .    
\end{equation}
Here $q_{3}, Q^{c}_{3}, \text{ and } D_{3}^{c}$ are third family
copies of the fields we have included for the first two families,
$D_{3}$ is a new field for the third family, and we do not include a
$U^{c}_{3}$ or $Q_{3}$ field.\footnote{This third family field content
  is anomalous. The anomalies can be canceled with additional fields
  with masses at the TeV scale. For example, adding the set
  $ \left\{ U^{c}_{3},D'^{c}_3,Q_{3} \right\}$ would do the
  trick.}  The top quark acquires a mass from the first term, while
the last term gives a large Dirac mass for the pair $D_{3}$ with the
lower component of $Q^{c}_{3}$. Consequently the bottom anti-quark is
mostly the $SU(2)_{M}$ singlet field $D^{c}_{3}$, and the mixing angle
for the $b$ quark is negligible.  Thus we have
\begin{equation}
  \label{eq:25}
  s_u=s_c\,,  \quad s_d=s_s, \quad  s_t=1\,, \quad s_b=0, \text{ and }
  s_e=s_\mu=s_\tau  \ . 
\end{equation}

\subsection{$SU(3)$ flavor model}

It would be especially attractive to accommodate the large top mass in
a fully $SU(3)$ flavor symmetric Lagrangian of the form of
\eqref{eq:23} including 3 copies of all the fields in Table
\ref{tab:1}. Our previous model treated the third family differently
in expectation of difficulty in obtaining a large top quark mass from
the small Higgs vev $v_{X}$, but it is worth exploring if a more
flavor symmetric Lagrangian is viable.

Generally speaking our precision fits prefer small values of the vev
$v_{X}$, as we assumed in our introductory section. However this would
necessitate a large Yukawa coupling in order to realize the large top
mass. To avoid potential problems with strong coupling (and to remain
within the validity of our perturbative analysis) we will limit the
size of the Yukawa coupling, which in turn requires a not-so-small vev
$v_{X}$. But a larger value of $v_{X}$ implies smaller bidoublet vevs,
which reduces the coupling of the $W'$ to $W Z$. To compensate for
this effect we are forced to larger values of the $SU(2)_{M}$
coupling, $g_{M}$. Consequently we need a compromise between large top
Yukawa and large $g_{M}$. The details of this compromise will be
explored in our precision fits.

There are then three relevant mixing angles for this $SU(3)$ flavor
symmetric model:
\begin{equation}
  \label{eq:26}
  s_u=s_c=s_{t},\quad s_d=s_s=s_{b}, \text{ and }     s_e=s_\mu=s_\tau \ .
\end{equation}

\section{Fit to precision electroweak and LHC data}
\label{sec:PEW}

Here we consider a simultaneous fit of our models to precision
electroweak data and the diboson signal. We also include bounds from
$W'$ decay to dijet resonance searches and from $Z'$ decay to dilepton
resonance searches. The fit confirms and validates our simplified
analysis in the Introduction.

The focus of our paper is the diboson signal and we therefore
constrain the parameters of our model to produce a fixed diboson cross
section at 8 TeV, $\sigma_{WZ}$. The remaining data is incorporated by
minimizing a global $\chi^2$ function
\begin{equation}
  \label{eq:27}
  \chi^2_\text{total}=\chi^2_{ll} +\chi_{jj}^2+\chi_{PEW}^2\ .  
\end{equation}
The values and choices we have made for each of these is detailed
below.

\begin{description}
\item[$\sigma_{WZ}$ diboson cross section:] In order to reproduce the
  observed diboson signal from \runi{} we fix the $W'$ mass to 1.9 TeV
  and the cross section times branching fraction to $W Z$ to
  $\sigma_{WZ}(8 \text{ TeV}) = 4 \text{ fb}$.  Values of $M_{W'}$
  within the range 1.8--2.0 TeV give similarly good fits to both the
  diboson data\cite{Dias:2015mhm} and the overall
  $\chi^2_\text{total}$. The best fit value for $\sigma_{WZ}$ from
  \runi{} is in tension with the 95\% confidence level upper bound
  obtained in \runii{} $\sigma_{WZ} < 25$ fb (see \cite{Dias:2015mhm,
    Bellomo_Moriond} for a summary of the \runii{} searches for
  $W'\rightarrow WZ$ and $W'\rightarrow Wh$ \cite{CMS-PAS-EXO-15-002,
    ATLAS-CONF-2015-068, ATLAS-CONF-2015-071, ATLAS-CONF-2015-073,
    ATLAS-CONF-2015-075, ATLAS-CONF-2015-074}). To translate this
  bound into an equivalent 8 TeV cross section bound we use a six-fold
  parton luminosity scaling from 8 TeV to 13 TeV, yielding our target
  value for the cross section of $4$ fb.
\item[$\chi^{2}_{ll}$ dileptons from $Z'$ decay:] In \runi{} both
  ATLAS\cite{Aad:2014cka} and CMS\cite{Khachatryan:2014fba} searched
  for the decay of a narrow resonance to dileptons.  In CMS no events
  were seen above 1.9 TeV. Combining the searches for dimuons and
  dielectrons with assumed lepton-flavor universality, CMS obtained a
  95\% confidence upper limit of 0.09 fb on the cross section times
  branching fraction to one species of dileptons. ATLAS saw no events
  above 2 TeV and obtained a bound of 0.2 fb for the same
  observable. Assuming Poisson statistics with zero observed events
  and combining the two bounds into a single \runi{} likelihood
  $\mathcal{L} = e^{-N_{CMS}}e^{-N_{ATLAS}}$ allows us to define an
  equivalent $\chi^2=-2\log\mathcal{L}$
  \begin{equation}
    \chi^2_{ll} = 2\, (N_{CMS} + N_{ATLAS}) = 6\,
    (\frac{\sigma_{ll}}{0.09 \text{ fb}}+ \frac{\sigma_{ll}}{0.2
      \text{ fb}}) = 100
    \text{ fb}^{-1} \,\sigma_{ll} \ .  
  \end{equation} 
  Here we used the fact that with Poisson statistics zero observed
  events gives a 95\% confidence bound on the number of expected
  events $N^{95\%}=-\log(0.05) \simeq 3$, irrespective of the number
  of expected background events.
\item[$\chi^{2}_{jj}$ dijet events:] \runi{} data from both CMS and
  ATLAS showed an intriguing $\sim 2 \sigma$ excess of dijet events
  with dijet invariant mass near 1.9 TeV. This data could arise from
  the $W'$ decaying to dijets with a cross section of
  $\sigma_{jj}(8 \text{ TeV}) \sim 50-100$
  fb\cite{Dobrescu:2015qna,Brehmer:2015cia,Dobrescu:2015yba,Dobrescu:2015jvn}.
  Unfortunately, neither CMS\cite{Khachatryan:2015dcf} nor
  ATLAS\cite{ATLAS:2015nsi} confirmed this excess in \runii{} and
  instead set bounds, with the stronger bound coming from ATLAS. In
  our model the $W'$ width is less than the energy resolution in
  ATLAS, and using an acceptance times efficiency of $\sim 50\%$ the
  limit is $\sigma_{jj}(13 \text{ TeV}) \lesssim 150$ fb.  Translating
  this bound into an equivalent 8 TeV bound by multiplying by first
  generation $\bar q q$ parton luminosity ratios we obtain
  $\sigma_{jj}(8 \text{ TeV}) \lesssim 24$~fb at 95\% confidence.  The
  fermion mixing angles that govern the coupling of the $W'$ to quarks
  $g_{ud} = s_u s_d g_M$ allow accommodation of this bound. However,
  reducing the dijet branching fraction of the $W'$ below that of the
  diboson branching fraction would require fine-tuning of parameters
  (see the discussion after eq.  \eqref{eq:18}).  In order to disfavor
  this fine-tuned region of parameter space and motivated by the
  preference for dijets from \runi{} we include a non-zero central
  value for the dijet rate in our fit
  $\sigma_{jj}(8 \text{ TeV})=12 \pm 6$ fb.  The uncertainty is chosen
  so that the $2\sigma$ upper bound coincides with the ATLAS 95\%
  confidence limit.  Thus we take
  \begin{equation} 
    \chi^2_{jj} = \left(
      \frac{\sigma_{jj} (8 \text{ TeV}) - 12 \text{ fb}}{6 \text{
          fb}}\right)^2
  \end{equation}
\item[$\chi^{2}_{PEW}$ precision electroweak observables:] We include
  all precision electroweak observables listed in the most current
  review of the Particle Data Group\cite{Agashe:2014kda}. Most
  important in this list are the masses, widths, and line shapes of
  the $W$ and $Z$, precision measurements of the fermion couplings in
  $Z$ decay branching fractions and forward-backward asymmetries. This
  fit is conveniently implemented by using the work of Han and
  Skiba\cite{Han:2004az} who combined all constraints from precision
  electroweak measurements into a single $\chi^2_{PEW}$.  We will
  describe this formalism and our modifications in the following.
\end{description}

We updated the precision electroweak function $\chi_{PEW}^2$ of
\cite{Han:2004az} to include the best fit Higgs mass and the latest
values for precision observables from the Particle Data
Group\cite{Agashe:2014kda}. The Han and Skiba $\chi_{PEW}^2$ function
depends on the coefficients of universal dimensions 6 operators
obtained by integrating out new physics heavier than the electroweak
scale.  Thus to apply the formalism to our model we integrate out the
$W'$ and $Z'$ and extract the coefficients of the dimension 6
operators so generated.  Since the three families of fermions have
different $W'$ and $Z'$ couplings in our models we generalize the
operator basis in \cite{Han:2004az} to allow for non-universal
operator coefficients (for similar such generalizations see
\cite{Han:2005pr,Efrati:2015eaa}).

In the notation of \cite{Han:2004az} the operator coefficients are
\begin{equation}
  \label{eq:28}
  a_h= - \frac{(2 g_{h})^2}{2
    M_{Z'}^2} + \frac{(g'/\sin\theta_M)^2}{2 M_{W'}^2}\, t^4\ , \quad
  a_{hf}=- \frac{g_h g_f}{M_{Z'}^2} \ , \quad a_{ff'} = - \frac{g_f
    g_{f'}}{M_{Z'}^2}\ , 
\end{equation}
where
\begin{equation}
  \label{eq:29}
  g_h\equiv\frac{g'}{2\sin\theta_M \cos\theta_M}
  \left(
    \cos^2\theta_M   -\sin^{2}\theta_{X}
  \right),
\end{equation}
the fermion couplings $g_f$ were defined in (\ref{eq:22}) in terms of
the charges $T^{3}_{M} \text{ and } Y$, and
$\sin \theta_{X}\equiv v_{X}/v$ is the fraction of the EW breaking vev
coming from the $H_{X}$ vev.  Note that the operators in
\cite{Han:2004az} are written in terms of right-handed fields for
$SU(2)_{SM}$ singlets, $u_{R}, d_{R}, e_{R}$, and with this convention
the charges for the SM fields are
\begin{center}
  \begin{tabular}{>{$}c<{$}|>{$}c<{$}>{$}c<{$}>{$}c<{$}>{$}c<{$}>{$}c<{$}}
    \text{fermion field}\  f \quad\ &\quad  q \quad\ &\quad u_R\quad\
    &\quad d_R\quad\ &\quad  l\quad\ &\quad e_R \\ 
    \hline{}
    T^3_M  & 0 & \frac12 & -\frac12 &0 &-\frac12 \\
    Y & \frac16 & \frac23&  -\frac13& -\frac12& -1 \\
  \end{tabular}
\end{center}
No other operators are generated at tree level.  We will find that the
fit prefers couplings of order 1 or smaller and loop-generated
operators can be neglected.  The operators generated from integrating
out the $Z'$ are easily recognized as they are proportional to
$1/M_{Z'}^2$.  These operators all involve contractions of
$SU(2)_{SM}$-singlet currents, and the triplet operators in the
Han-Skiba basis have vanishing coefficients in our model. Integrating
out the $W'$ generates a contribution to $a_h$ (\textit{i.e.\/} the
$T$-operator) as already discussed in the Introduction. All other
dimension six operators mediated by the $W'$ are unimportant for
several reasons: {\it i.} leptonic operators involve the right-handed
neutrinos which we assume to be too heavy to be relevant to precision
physics, {\it ii.} operators with only quarks are not sufficiently
well constrained by data, and {\it iii.}  operators which lead to
effective couplings of right-handed fermions to the $W$ do not have an
SM counterpart to interfere with. Therefore their contributions to
observables are as small as contributions from dimension 8 operators
which we have consistently ignored.

In addition to the usual SM couplings and the $W'$ mass (fixed to
$M_{W'}=1.9$ TeV) both our models have the following continuous free
parameters
\begin{equation}
  \label{eq:30}
  \cos\theta_{M}, t^{2}, s_{u}, s_{d}, s_{e}, \sin \theta_{X}
\end{equation}
Fits for the $SU(2)$ flavor model will prefer very small values of
$\sin \theta_{X}$ and consequently this angle plays little role in our
analysis of this model. In the $SU(3)$ model the fits also prefer
small values of $\sin \theta_X$. However, in this case the top quark
mass is given by $m_t=y_t c_u v \sin\theta_X/\sqrt{2}$ so that small
$\sin \theta_X$ requires large $y_t$ to compensate.  In order to
remain safely in the perturbative part of parameter space we impose
the constraint $y_t\leq 2$. This limits the size of
$\sin\theta_{X} \ge m_t /( c_u v \sqrt{2})$.  Since the fit prefers
small values of $v_{X}$, the best fit point is always near the
smallest possible value for $\sin\theta_{X}$.  We can therefore
simplify our analysis by fixing $\sin\theta_{X}$ in our fits for the
$SU(3)$ model to saturate this inequality:
$\sin \theta_X= m_t /( c_u v \sqrt{2})$.

The choice of $SU(2)_M$ representation for the scalar field $H_M$
introduces an additional discrete parameter
$k = 1, \sqrt{2}, \sqrt{3}, \ldots$ that enters the relationship
between the $Z'$ and $W'$ masses $M_{Z'}=k M_{W'}/\cos\theta_M$. We
focus on the two simplest cases: doublet breaking with $k=1$ and
triplet breaking with $k=\sqrt{2}$. Larger representations for
$H_{XM}$ would lead to heavier $Z'$ masses for which good fits to the
precision electroweak data can also be obtained.

Table \ref{tab:fitresults} shows the best fit parameters for the two
Models and the two choices $k=1$ and $k=\sqrt{2}$. For each case we
show the mass of the $Z'$, the expected $W'$ to dijet and $Z'$ to
dilepton rates at 13 TeV, and two different measures of the goodness
of fit. The first measure is the difference between $\chi_{PEW}^2$ of
the best fit point relative to the SM,
$\triangle \chi^2_{PEW} \equiv~\left.\chi^2_{PEW}\right|_\text{best
  fit}-\left.\chi^2_{PEW}\right|_{SM}$, while the second is the
difference of the overall $\chi^2_\text{total}$ relative to the SM,
$\triangle \chi^2_\text{total} \equiv
\left.\chi^2_\text{total}\right|_\text{best
  fit}-\left.\chi^2_\text{total}\right|_{SM}$.  Note that differences
in $\chi^2_\text{total}$ on the order of a few should be taken with a
grain of salt because of the somewhat arbitrary choice of central
value for the dijet cross section in $\chi^2_{jj}$.
\begin{table}[t]
  \centering
  \begin{tabular}{>{$}c<{$}|>{$}c<{$}|>{$}c<{$}|>{$}c<{$}|>{$}c<{$}}
    &  SU(2) \text{Model} & SU(3) \text{Model} & SU(2)  \text{Model}& SU(3)  \text{Model} \\
    &  k=\sqrt{2} & k=\sqrt{2} &  k=1 & k=1  \\
    \hline{}
    \cos\theta_M  & 0.92& 0.97 & 0.87 &0.94\\
    t^2 & 0.50& 0.40& 0.62 &0.55 \\
    \langle H_X \rangle\ [ \text{GeV}]& 0& 137& 0 &127 \\
    s_d & 0.32 & 0.20& 0.37 & 0.34 \\
    s_e & 0.30 & 0.29& 0.56 & 0.39  \\
    s_u  & 0.54 & 0.54&  0.57 & 0.42 \\
    \hline{}
    M_{Z'}\, [ \text{TeV}]& 2.92 & 2.78& 2.18 & 2.02  \\ 
    \sigma_{jj}(13 \text{ TeV})\, [ \text{fb}]& 74 & 66& 71 & 62  \\ 
    \sigma_{ll}(13 \text{ TeV})\, [ \text{fb}]& 0.03 & 0.08& 0.17 & 0.11  \\ 
    \hline{}
    \triangle\chi^2_{PEW} & -1.1 & -1.7& 1.1 & 0.5  \\
    \triangle\chi^2_\text{total} & -4.9 & -5.1& -0.8 & -2.0  \\
  \end{tabular}
  \caption{Best fit points and predictions in the 4 Models for fixed
    $\sigma_{WZ}(8 \text{ TeV})=4\text{ fb}$ corresponding to
    $\sigma_{WZ}(13 \text{ TeV})\simeq 24\text{ fb}$.  Note that
    $\sin\theta_X=\langle H_X\rangle/v$.}
  \label{tab:fitresults}
\end{table}

As expected the best fit points for models with $k=\sqrt{2}$ have $Z'$
masses larger than those for models with $k=1$, and therefore more
easily avoid constraints from both direct searches for
$Z'\rightarrow l\bar l$ in \runi{} and \runii{} and precision electroweak
measurements. We further explore models with $k=\sqrt{2}$ in the next
subsection. While the models with $k=1$ have a significantly worse
$\triangle\chi^{2}$ compared to the $k=\sqrt{2}$ models, a small
region of parameter space which satisfies all constraints exists. We
discuss this case following the $k=\sqrt{2}$ analysis.

\subsection{Models with triplet breaking  $k=\sqrt{2}$}

\begin{figure}[htb]
  \begin{center}
    \includegraphics[width=.4\textwidth]{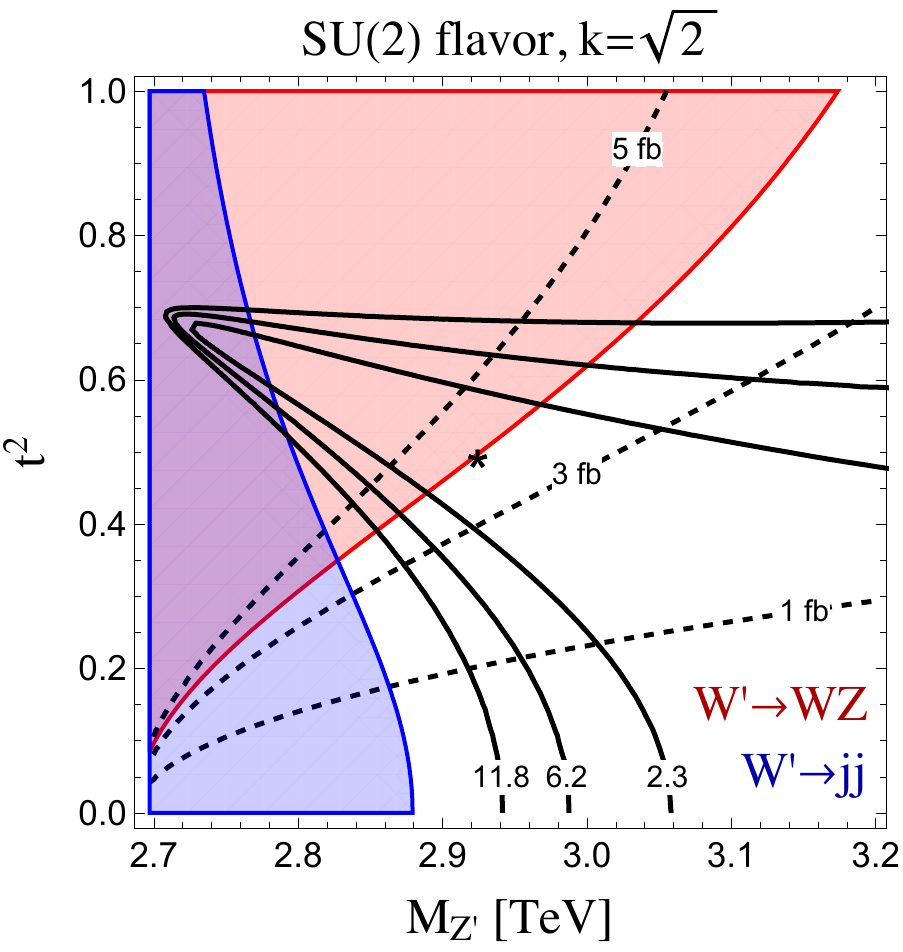}
    \qquad\qquad{}
    \includegraphics[width=.4\textwidth]{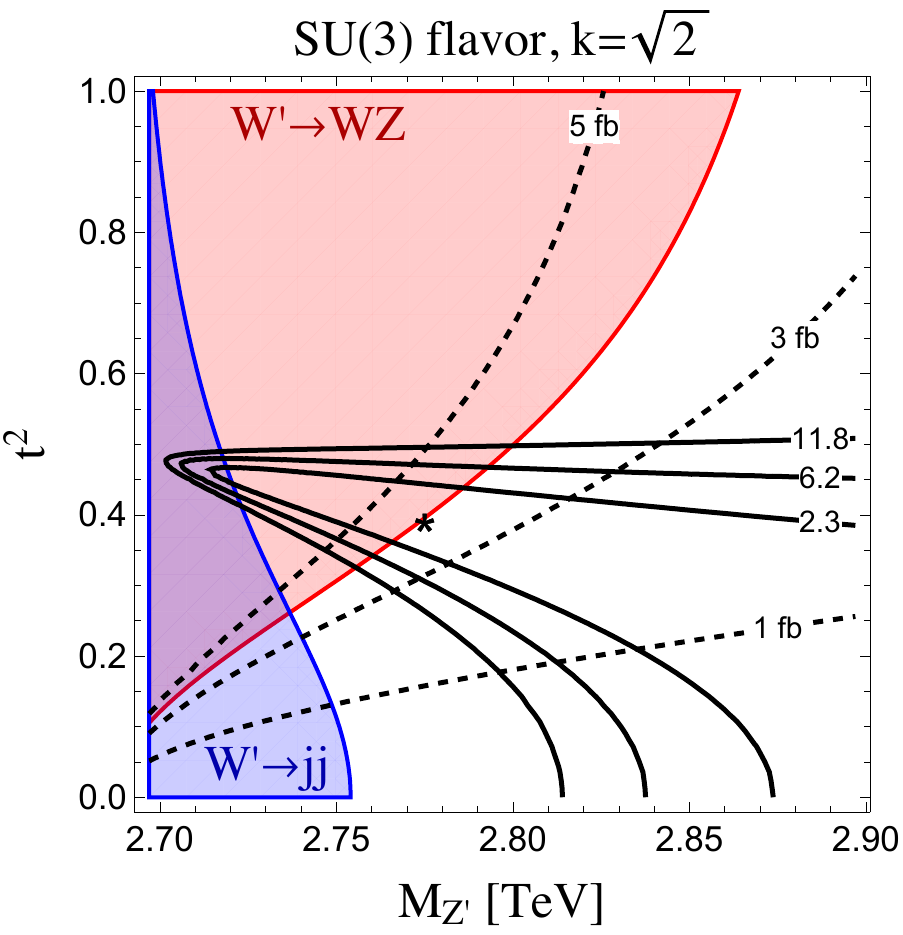}
  \end{center}
  \caption{Preferred and excluded regions in the $M_{Z'}$ versus $t^2$
    parameter space in $SU(2)$ and $SU(3)$ flavor models with
    $k=\sqrt{2}$. The fermion mixing parameters $s_e, s_d, s_u$ are
    fixed at the best fit values in Table \ref{tab:fitresults}. Solid
    lines are contours of constant $\triangle \chi^2_{PEW}$, enclosing
    68\%, 95\%, and 99.7\% confidence regions in the Gaussian
    approximation. Dashed lines are contours of constant 8~TeV $W Z$
    diboson cross section. Colored regions are excluded by 
    diboson searches from ATLAS and CMS at 13 TeV and dijet resonance
    searches from ATLAS and CMS at 13 TeV.  The best fit points of
    Table \ref{tab:fitresults} are indicated by asterisks.  A
    satisfactory PEW fit with a sizable diboson cross section (for
    example $\sigma_{WZ} \gtrsim 3$ fb) fixes the $Z'$ mass to lie
    near 3.0~TeV in the $SU(2)$ model and
    near 2.8~TeV in the $SU(3)$ model.}
  \label{fig:mZpvst2_plots}
\end{figure}

Both models with $k=\sqrt{2}$ allow excellent fits, obtaining the
diboson and dijet signals while avoiding constraints from
$Z'\rightarrow l\bar l$ searches.  Both also have precision electroweak
fits that improve upon the SM. In all cases the $Z'$ mass is predicted
to be near the range 2.8--3.0 TeV and out of reach of the \runi{}
dilepton search for generic values of the fermion mixing angles.

To understand the robustness of the fits and explore the parameter
spaces of the two $k=\sqrt{2}$ models we plot the main LHC observables
and $\chi^2_{PEW}$ as a function of the model parameters
$M_{Z'}=\sqrt{2}/\cos\theta_M \times 1.9$ TeV and $t^2$ in Figure
\ref{fig:mZpvst2_plots}.  In these plots we hold the remaining
parameters fixed to their best fit values shown in Table
\ref{tab:fitresults}.
The colored regions in the plots correspond to the direct $95\%$
confidence search limits from \runi{} and \runii{} at the LHC.  The
$Z' \rightarrow l\bar l$ searches place no restrictions on the parameter
space shown. However, both dijet and diboson searches exclude
significant portions of this parameter space.  We also plot contours
of constant $\triangle \chi^2_{PEW}$ relative to the point which
minimizes $\chi^2_{PEW}$.  In the Gaussian approximation the contours
labeled $2.3, 6.2, 11.8$ then correspond to $68\%, 95\%, 99.7\%$
confidence regions in this two-dimensional parameter space.

For each model a large region of parameter space satisfies both
precision electroweak constraints and direct searches.  Requiring a
sizeable 8 TeV diboson signal narrows the allowed region to a small
domain near the best fit point. Within this domain the $Z'$ mass is
predicted to lie near 2.8--3.0 TeV.

Not visible in these plots is a somewhat flat direction for
$\chi^2_\text{total}$ along the axis of the wedge-shaped region
bounded by the precision electroweak contours. In moving along this
trough in parameter space the hidden parameters $s_d, s_u$ can be
adjusted to avoid the constraints from diboson and dijet searches. For
smaller values of $M_{Z'}$ the trough gets increasingly narrow. This
is a sign that $t^2$ must be finely tuned to maintain a good PEW
fit. For larger values of $M_{Z'}\gtrsim 3.1$ TeV the combined
requirement of a sizeable diboson signal with a good PEW fit can no
longer be satisfied. Thus the $Z'$ mass prediction is quite robust,
with both models requiring a $Z'$ mass between 2.7 and 3.1 TeV.

This prediction is intriguing in light of a di-electron event with
invariant mass of $\simeq 2.9$ TeV observed by CMS in
\runii{}\cite{CMS-PAS-EXO-15-005}. The likelihood that this event is
due to SM backgrounds is quite small: these backgrounds contribute
only $0.036 \pm 0.009$ events integrated over all invariant mass
greater than 2.8 TeV\cite{CMS-PAS-EXO-15-005}. It is therefore
worthwhile asking whether our predicted 13 TeV cross section for $Z'$
production with subsequent $Z'\rightarrow ee$ decay makes this process
a likely explanation of the CMS event.

The predicted number of dilepton events in our model is most sensitive
to the parameters $s_u$ and $\cos\theta_M$ which determine the $Z'$
coupling to up-quarks.
\begin{figure}[h]
  \begin{center}
   \includegraphics[width=.4\textwidth]{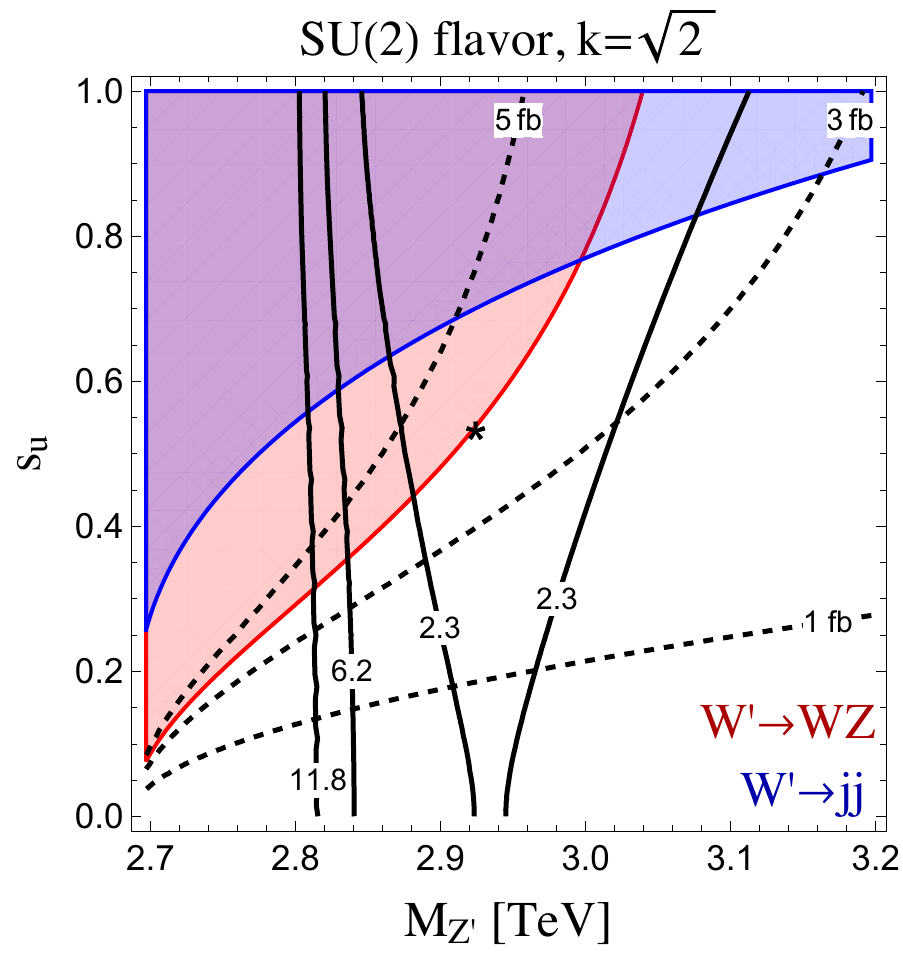}
    \qquad\qquad{}
    \includegraphics[width=.4\textwidth]{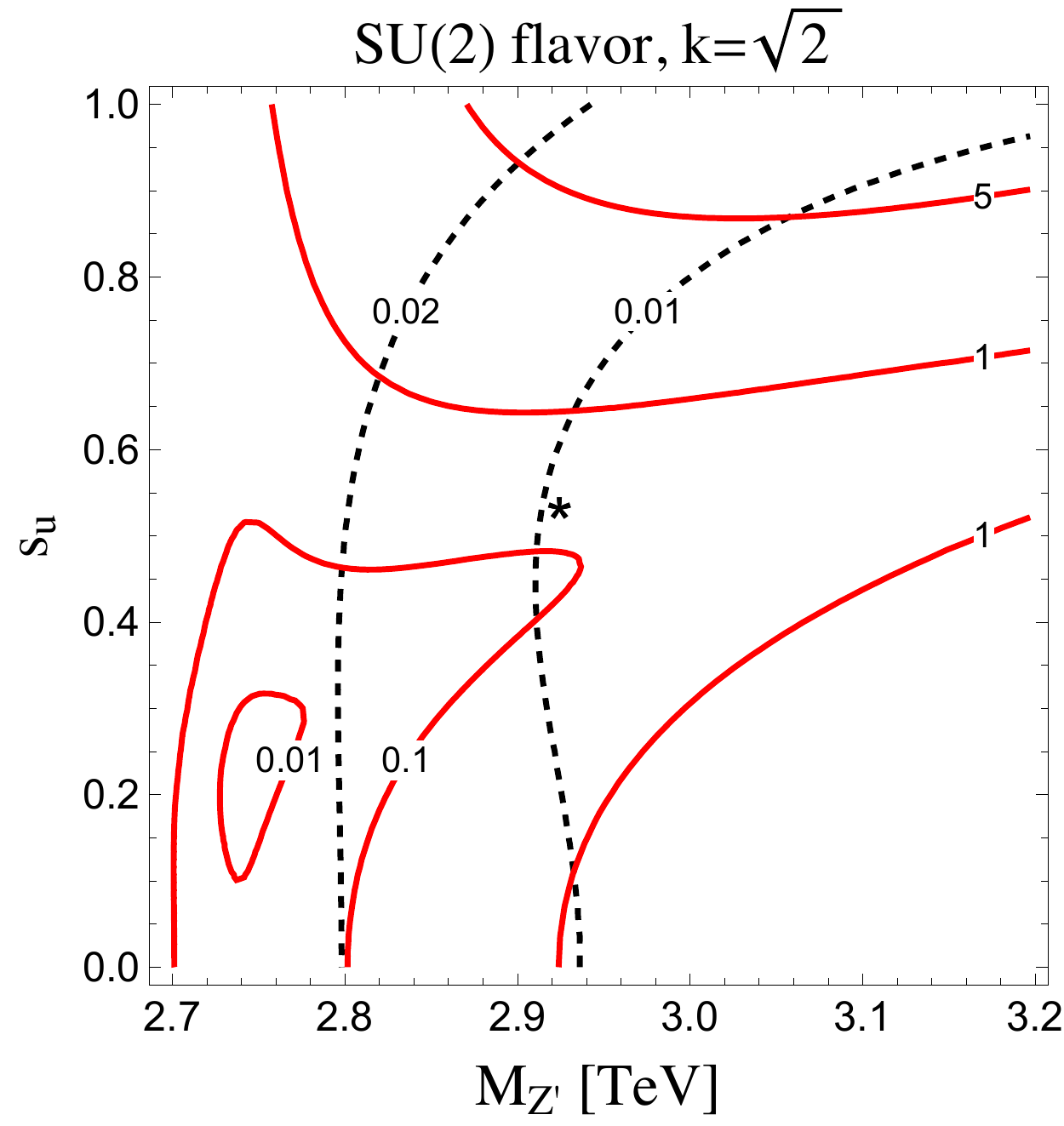}
  \end{center}
  \caption{Preferred and excluded regions in the $M_{Z'}$ versus $s_u$
    plane in the $SU(2)$ model
    with $k=\sqrt{2}$. The parameters
    $s_e, s_d, t^2$ are held fixed at the best fit values given in
    Table \ref{tab:fitresults}. The contours in the left plot are are
    as in Figure \ref{fig:mZpvst2_plots}. The plot on the right shows
    contours of the predicted number of dilepton $Z'\rightarrow l\bar l$
    events in \runii{} at CMS and ATLAS and electron plus muon final states
    combined (red, solid). Also
    shown are contours of constant $Z'$ width over mass
    (dashed). Note that in the parameter region with a satisfactory PEW
    fit and sizeable $W'\rightarrow WZ$ cross section the $Z'$ width
    is about 1\% and the predicted \runii{} dilepton event rate
    ranges from less than 0.1 events to 5 events.}
  \label{fig:mZpvssu_su2}
\end{figure}
\begin{figure}[h]
  \begin{center}
    \includegraphics[width=.4\textwidth]{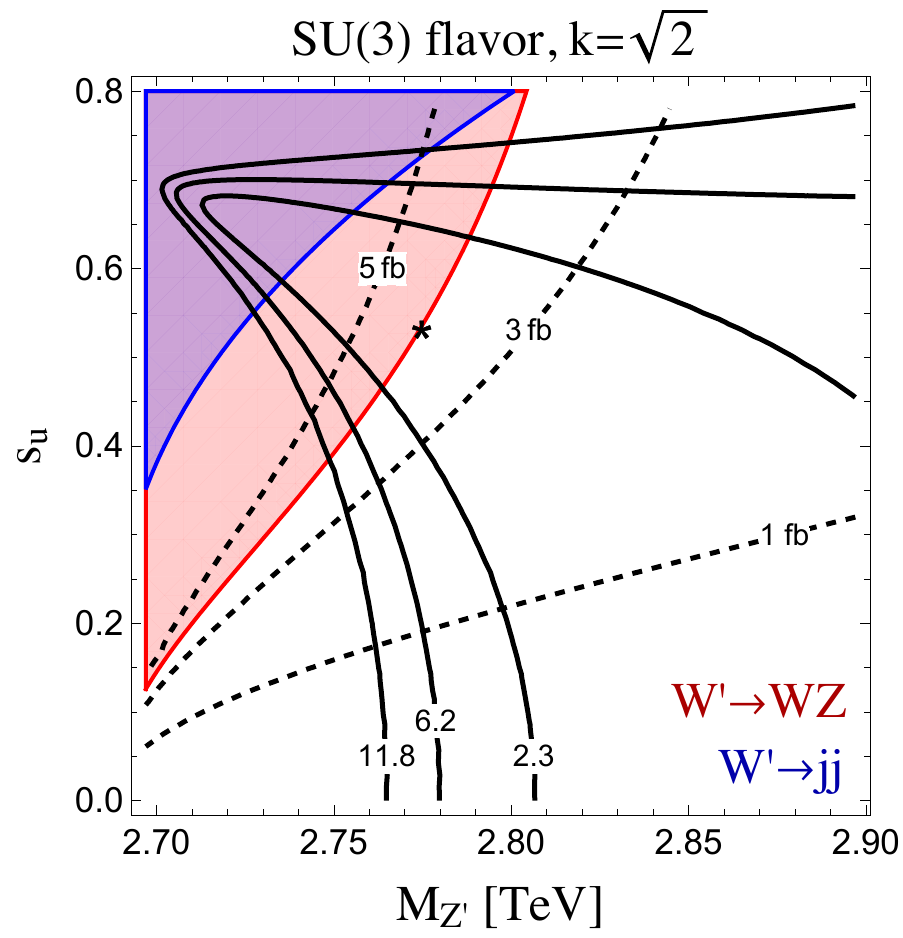}
    \qquad\qquad{}
    \includegraphics[width=.4\textwidth]{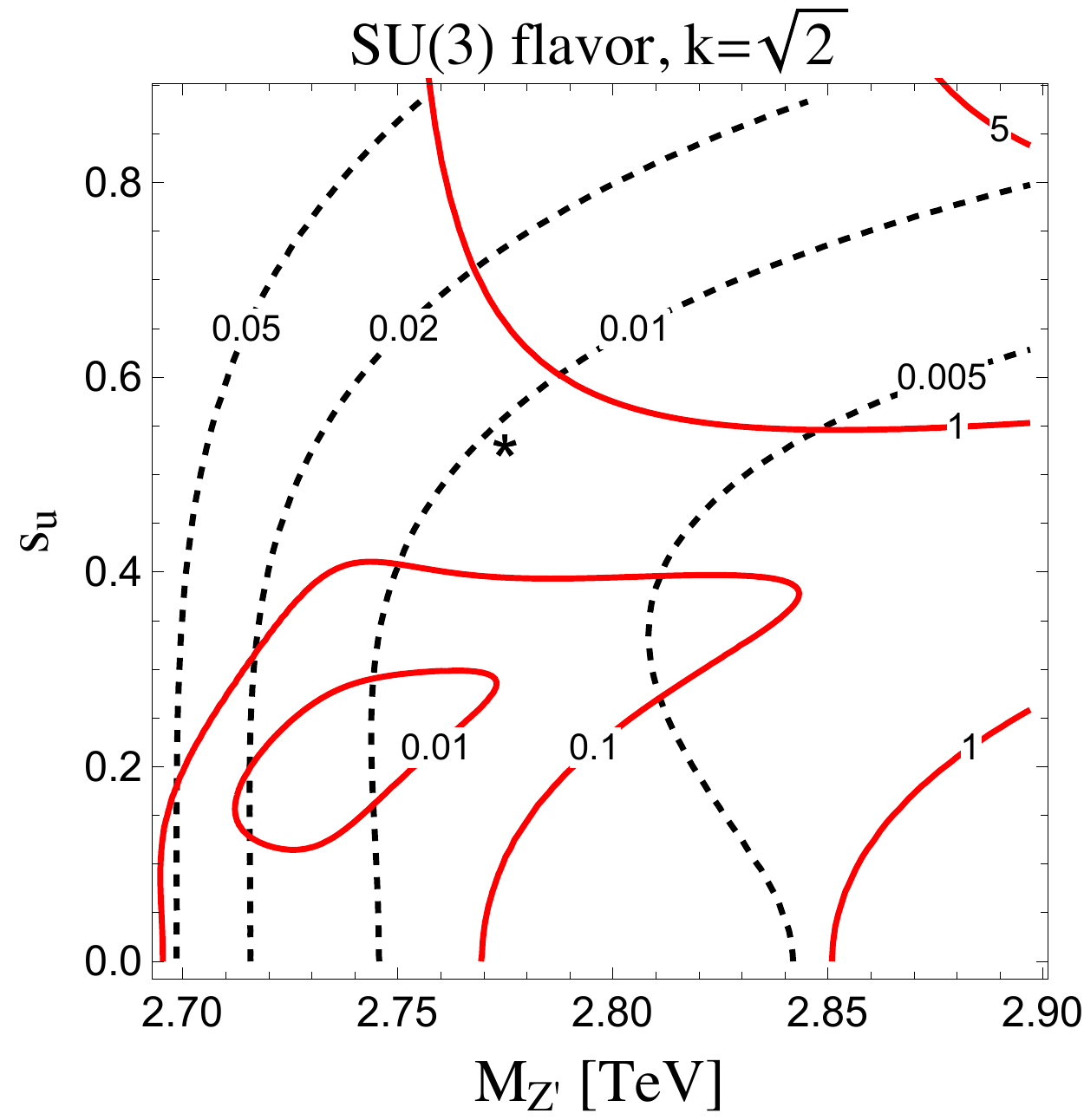}
  \end{center}
  \caption{Preferred regions of $M_{Z'}$ versus $s_u$ parameter space in the $SU(3)$
   model with predictions for the width of the $Z'$ and dilepton event rate at \runii{}.
   See caption for Figure \ref{fig:mZpvssu_su2} for details.}
  \label{fig:mZpvssu_su3}
\end{figure}
We therefore show the $Z'$ relative width $\Gamma/M$ (dashed) and
predicted dilepton event rates (red, solid) from $Z'\rightarrow l\bar l$
decay at \runii{} of ATLAS and CMS combined as a function of these
parameters in the right panels of Figures \ref{fig:mZpvssu_su2} and
\ref{fig:mZpvssu_su3}.  One sees that as $M_{Z'}$ decreases the gauge
coupling $g_M$ grows and the $Z'$ width increases. Similarly, large
$s_u$ implies a larger decay rate to up quarks and an increased width.
The solid red lines in the panel on the right indicate contours of
constant total number of $Z'\rightarrow l\bar l$ events predicted for CMS
and ATLAS and muons and electrons combined.  For the $SU(2)$ model in
the preferred region of parameter space near $M_{Z'} \simeq 2.9$ TeV,
the $Z'$ has a width of about 1\% and the number of events expected at
13 TeV varies between 0.05 and about 5. For the $SU(3)$ model, the
preferred $Z'$ mass is $M_{Z'}\sim 2.8$~TeV, and the width is also
about 1\% with between 0.1 and 2 events expected.  Both models are
perfectly consistent with the 1 observed $Z'\rightarrow e^+ e^-$ event
observed by CMS in \runii{} and promise many more events in the
upcoming 13 TeV runs.

\subsection{Models with doublet breaking $k=1$}

\begin{figure}[htb]
  \begin{center}
   \includegraphics[width=.4\textwidth]{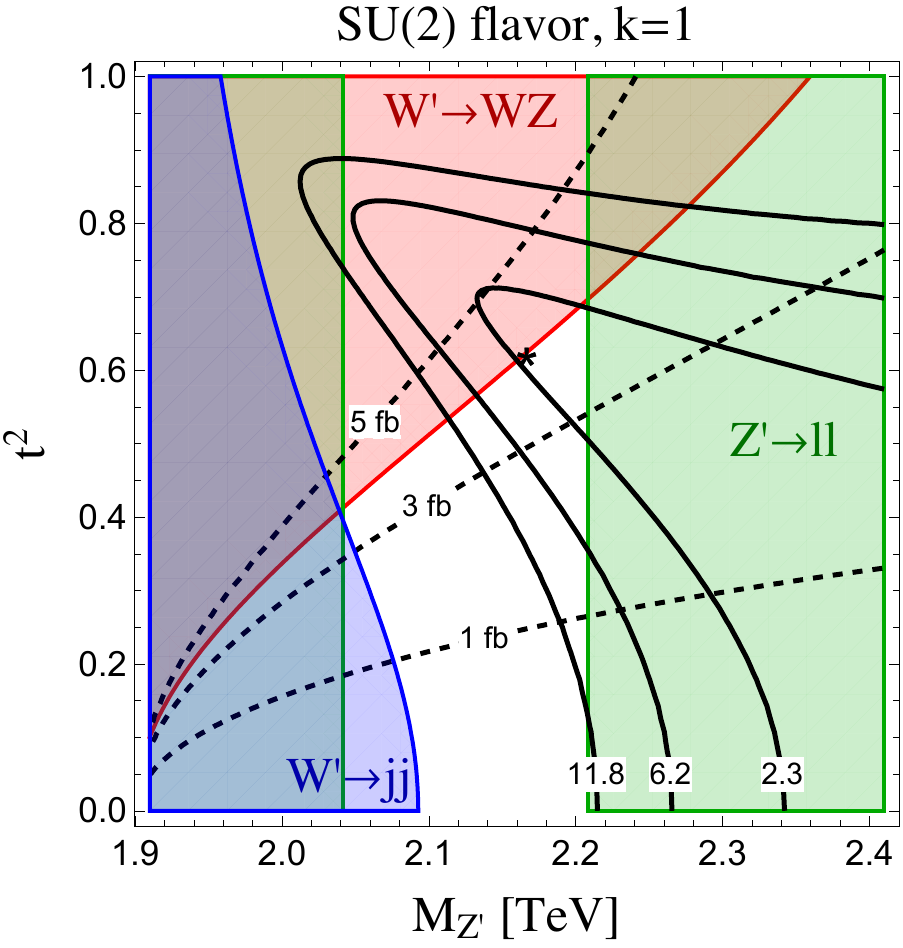}
    \qquad\qquad
   \includegraphics[width=.4\textwidth]{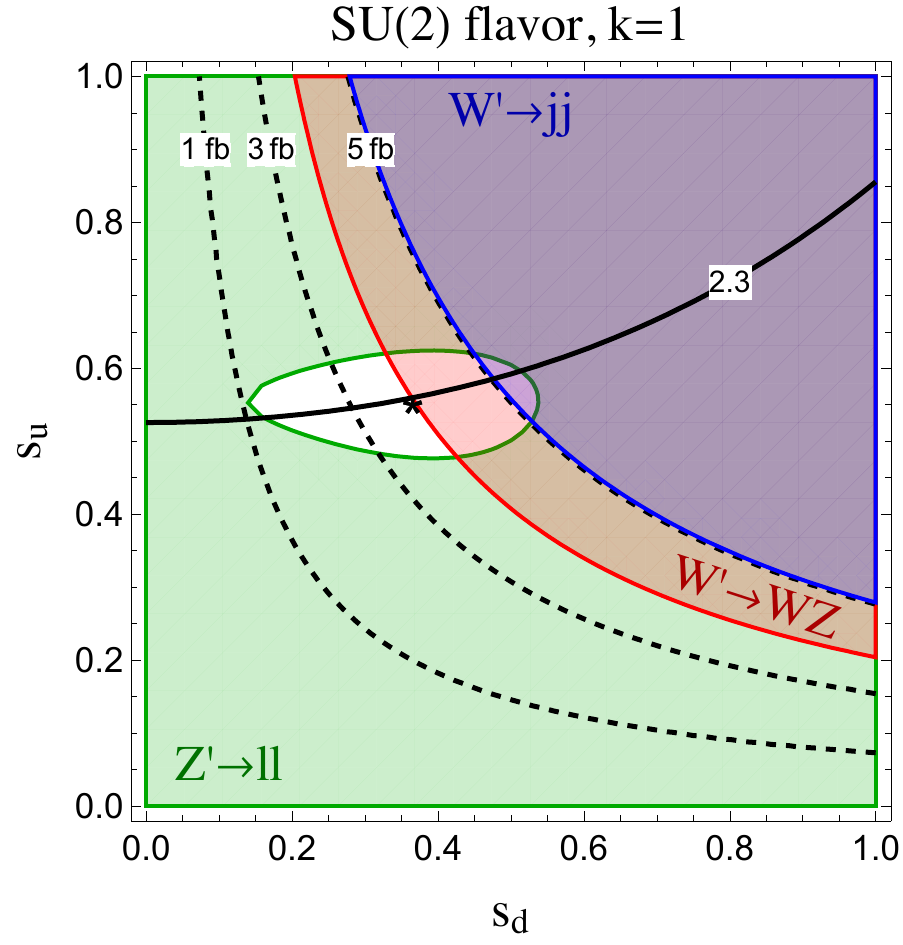}\\ \vskip .3cm
   \includegraphics[width=.4\textwidth]{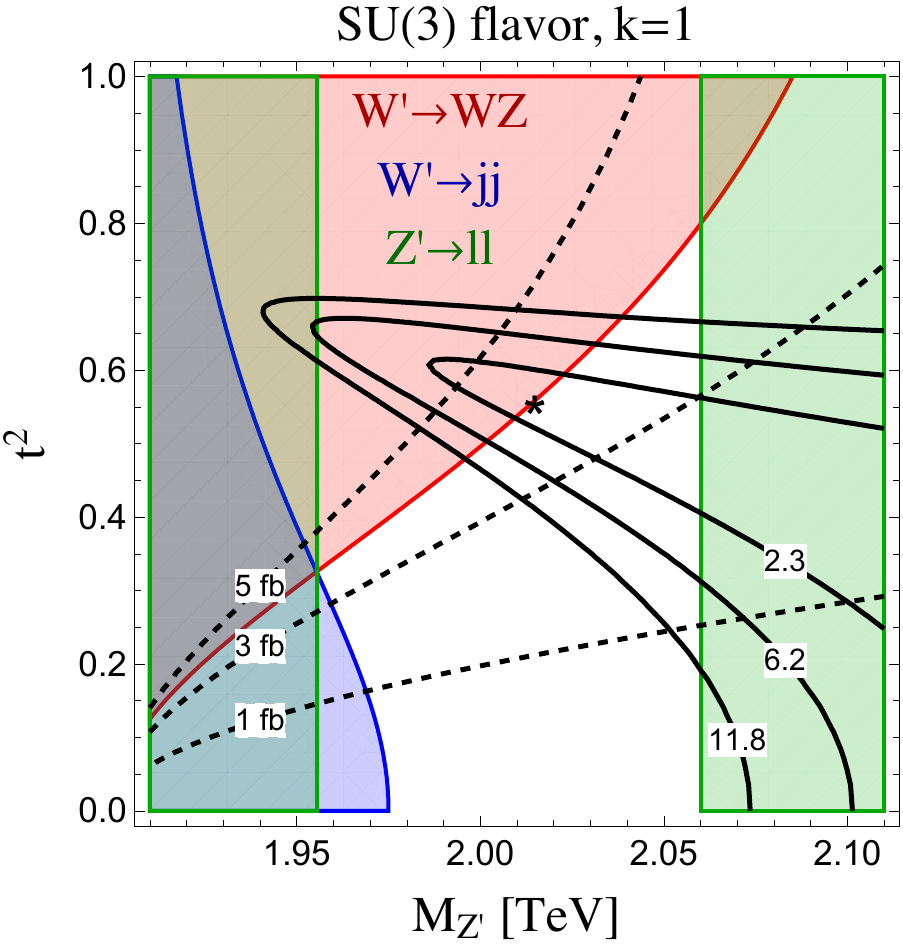}
    \qquad\qquad
   \includegraphics[width=.4\textwidth]{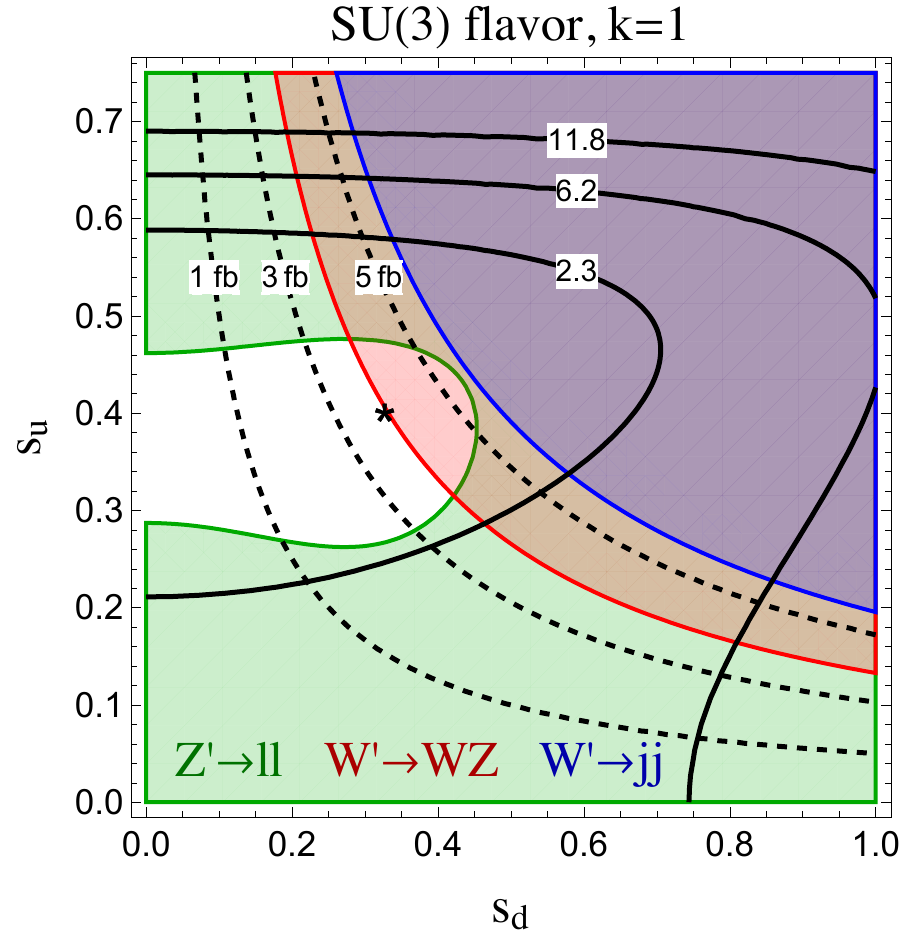}
  \end{center}
  \caption{\label{fig:model2k1} Preferred and excluded regions in the
    $M_{Z'}$ versus $t^2$ and $s_d$ versus $s_u$ parameter space in
    the $SU(2)$ and $SU(3)$ models with $k=1$. Note that most of
    parameter space is ruled out by the dilepton searches at ATLAS and
    CMS at 8 TeV and 13 TeV combined (green). The remaining allowed
    region has finely tuned values for $s_u$ and $s_d$ which minimize
    the $Z'$ production cross section.}
  \label{fig:plots_21Zt}
\end{figure}
The $SU(2)$ and $SU(3)$ models with $k=1$ predict a relatively light
$Z'$.  In both models it is possible to tune $s_u$ and $s_d$ such that
the $Z'$ coupling to quarks is very small (see Eq.~(\ref{eq:22})). In this
somewhat tuned region of parameter space the $Z'$ production cross section
sufficiently small to evade any $Z'$ search bounds.
In the $SU(2)$ model the viable region corresponds to a $Z'$ mass near 2.2 TeV
with a width well below 1\%. In the $SU(3)$ model there is a slightly
larger allowed region with $Z'$ mass near 2.0~TeV and also a very
narrow width.

Since neither \runi{} nor \runii{} have observed dilepton events at
2.0 or 2.2 TeV we can combine the dilepton bounds from ATLAS and CMS
at 8 and 13 TeV for both muons and electrons.\footnote{While there are
  some events---consistent with the tail of the Drell-Yan
  distribution---at 1.8 and 1.9 TeV, our $Z'$ is always heavier than
  $M_{W'}=1.9 \text{ TeV}$ and very narrow in the allowed parameter
  space. Therefore we discount the possibility that these events arise
  from $Z'$ production.}

\begin{figure}[htb]
  \begin{center}
    \includegraphics[width=.4\textwidth]{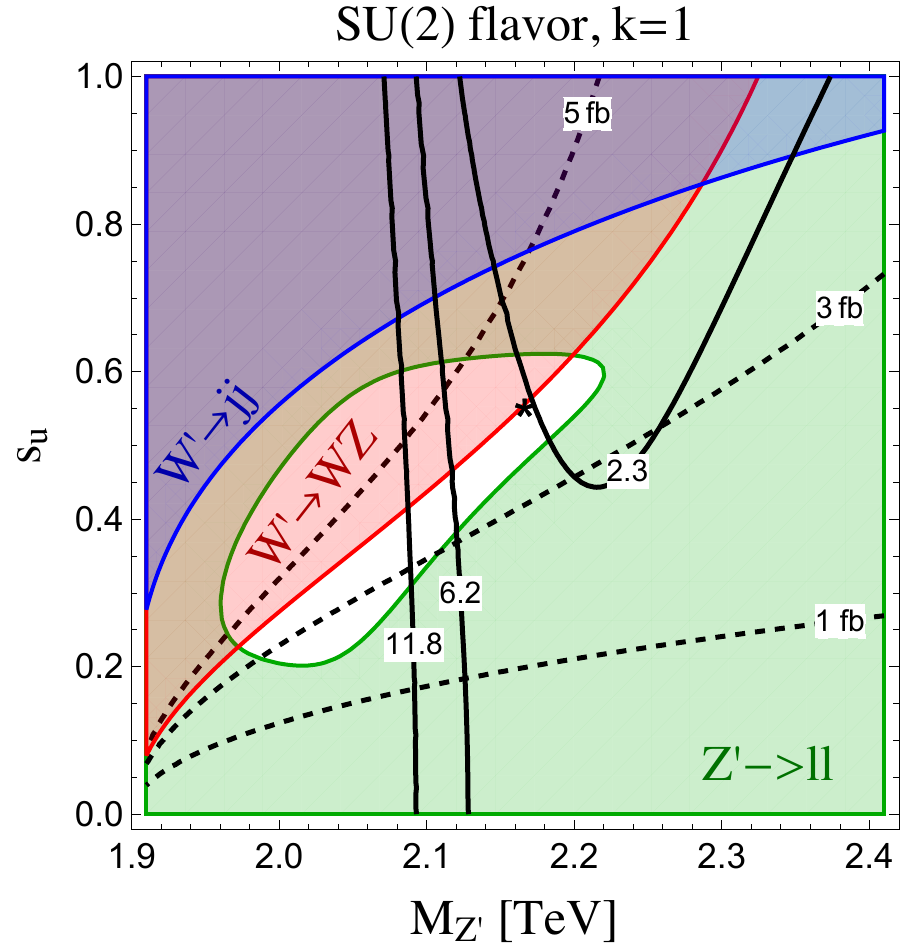}
    \qquad\qquad{}
    \includegraphics[width=.4\textwidth]{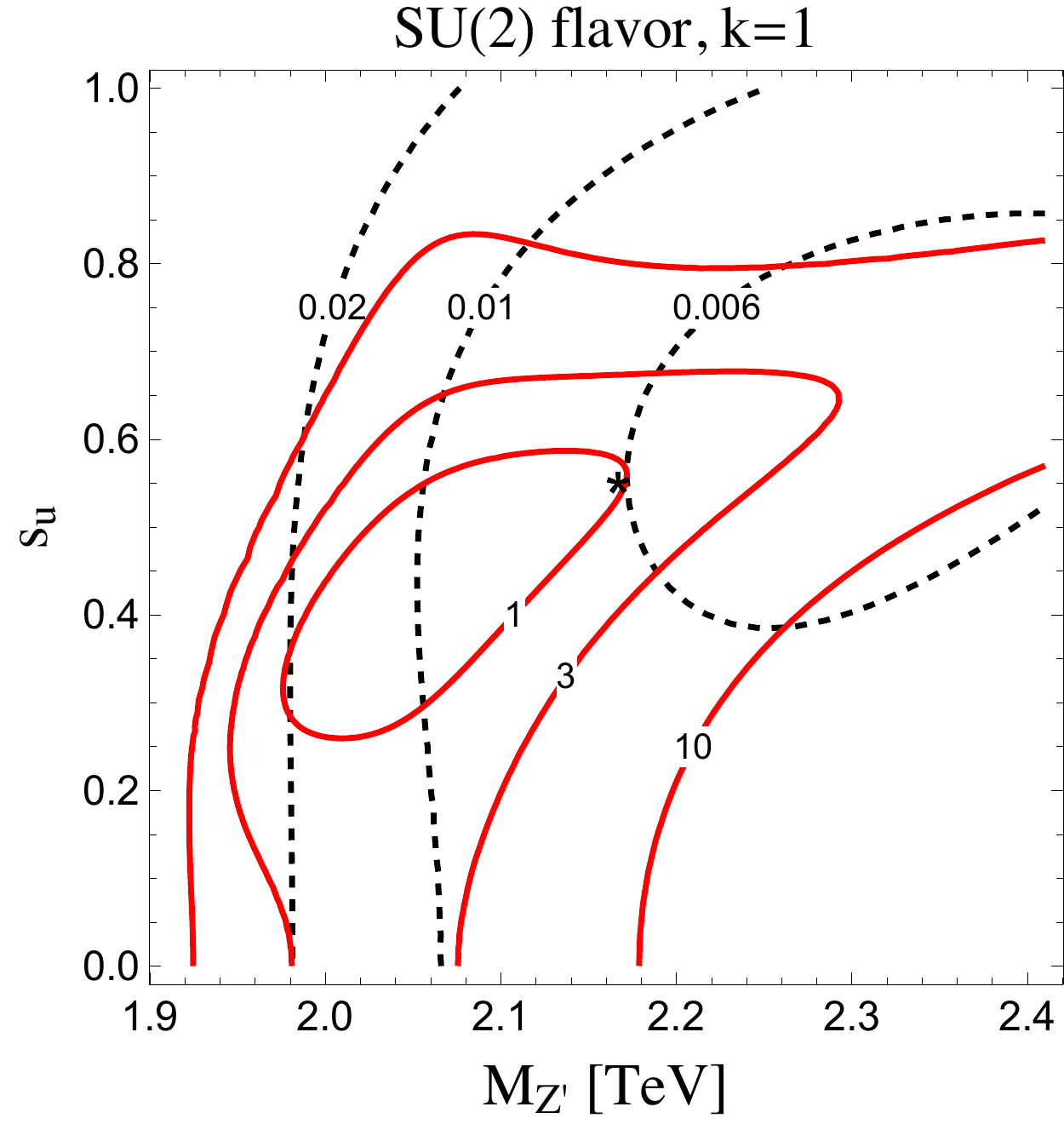}\\ \vskip .3cm
    \includegraphics[width=.4\textwidth]{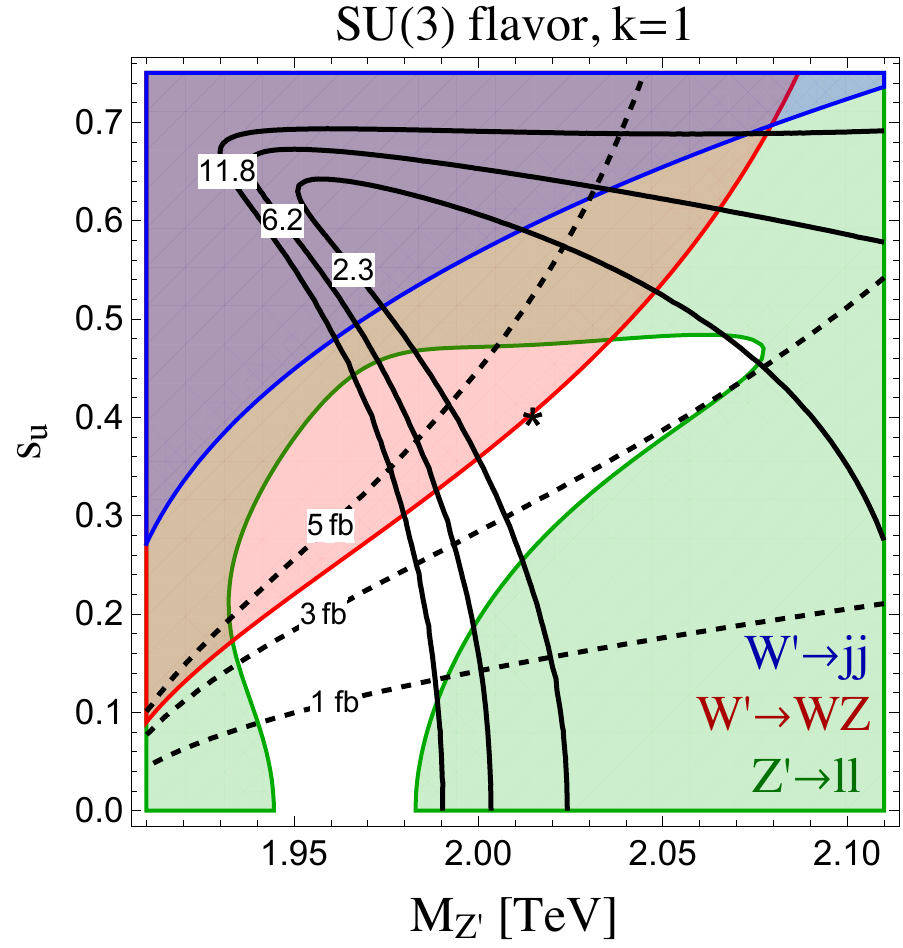}
    \qquad\qquad{}
    \includegraphics[width=.4\textwidth]{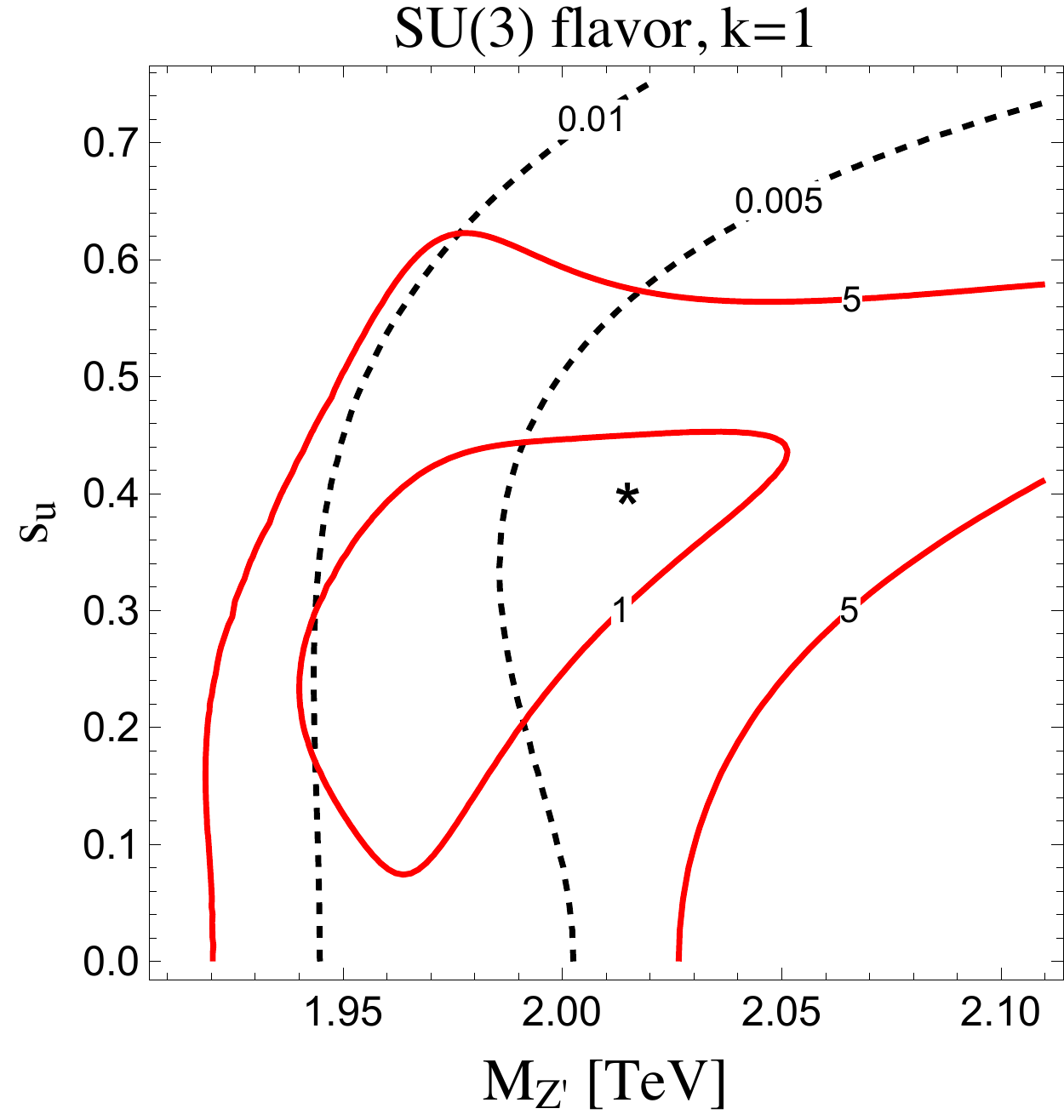}
  \end{center}
  \caption{\label{fig:model2k1contours} The plots on the left show PEW
    and direct search constraints in the $M_{Z'}$ \textit{vs.} $s_u$ parameter
    spaces of the $SU(2)$ and $SU(3)$ models for k=1.  The contour
    plots on the right show the predicted number of dilepton events at
    ATLAS and CMS combined for \runii{} (red, solid) and the predicted
    Z' width (black, dashed) in the same parameter space.}
  \label{fig:plots_21Zu}
\end{figure}

In Figure \ref{fig:plots_21Zt} 
we show two slices of parameter space. The plot in the right panel
shows that the viable region requires significant fine tuning of both
$s_u$ and $s_d$ to simultaneously avoid the dilepton bounds and obtain
an interesting $W Z$ diboson signal.  Figure \ref{fig:plots_21Zu}
shows the $Z'$ width and the expected \runii{} dilepton event rate
along side the allowed parameter spaces in the $M_{Z'}-s_u$ plane for
the two models.  Since the $Z'$ is very narrow in both cases and the
dilepton event rate is already very close to the 95\% confidence
bound, both $k=1$ models will be discovered or ruled out with only a
little additional 13 TeV running.

\medskip

\acknowledgments

We thank David E. Kaplan for useful discussions.  This work was supported by the
  U.S.\ Department of Energy's Office of Science.


\pagebreak

\phantomsection
\addcontentsline{toc}{section}{References}

\bibliographystyle{jhep}
\bibliography{dib_final_v2}

\providecommand{\href}[2]{#2}\begingroup\raggedright\begin{thebibliography}{10}

\bibitem{Aad:2015owa}
{\bf ATLAS} Collaboration, G.~Aad et~al., {\it {Search for high-mass diboson
  resonances with boson-tagged jets in proton-proton collisions at $\sqrt{s} =
  8$ TeV with the ATLAS detector}},
  \href{http://arxiv.org/abs/1506.00962}{{\tt arXiv:1506.00962}}.

\bibitem{Aad:2014xka}
{\bf ATLAS} Collaboration, G.~Aad et~al., {\it {Search for resonant diboson
  production in the $\mathrm {\ell \ell }q\bar{q}$ final state in $pp$
  collisions at $\sqrt{s} = 8$ TeV with the ATLAS detector}},  {\em Eur. Phys.
  J.} {\bf C75} (2015) 69, [\href{http://arxiv.org/abs/1409.6190}{{\tt
  arXiv:1409.6190}}].

\bibitem{Aad:2015ufa}
{\bf ATLAS} Collaboration, G.~Aad et~al., {\it {Search for production of
  $WW/WZ$ resonances decaying to a lepton, neutrino and jets in $pp$ collisions
  at $\sqrt{s}=8$ TeV with the ATLAS detector}},  {\em Eur. Phys. J.} {\bf C75}
  (2015), no.~5 209, [\href{http://arxiv.org/abs/1503.04677}{{\tt
  arXiv:1503.04677}}]. [Erratum: Eur. Phys. J.C75,370(2015)].

\bibitem{Aad:2015yza}
{\bf ATLAS} Collaboration, G.~Aad et~al., {\it {Search for a new resonance
  decaying to a W or Z boson and a Higgs boson in the $\ell \ell / \ell \nu /
  \nu \nu + b \bar{b}$ final states with the ATLAS detector}},  {\em Eur. Phys.
  J.} {\bf C75} (2015), no.~6 263, [\href{http://arxiv.org/abs/1503.08089}{{\tt
  arXiv:1503.08089}}].

\bibitem{Aad:2014aqa}
{\bf ATLAS} Collaboration, G.~Aad et~al., {\it {Search for new phenomena in the
  dijet mass distribution using $p-p$ collision data at $\sqrt{s}=8$ TeV with
  the ATLAS detector}},  {\em Phys. Rev.} {\bf D91} (2015), no.~5 052007,
  [\href{http://arxiv.org/abs/1407.1376}{{\tt arXiv:1407.1376}}].

\bibitem{Khachatryan:2014hpa}
{\bf CMS} Collaboration, V.~Khachatryan et~al., {\it {Search for massive
  resonances in dijet systems containing jets tagged as W or Z boson decays in
  pp collisions at $ \sqrt{s} $ = 8 TeV}},  {\em JHEP} {\bf 08} (2014) 173,
  [\href{http://arxiv.org/abs/1405.1994}{{\tt arXiv:1405.1994}}].

\bibitem{Khachatryan:2014gha}
{\bf CMS} Collaboration, V.~Khachatryan et~al., {\it {Search for massive
  resonances decaying into pairs of boosted bosons in semi-leptonic final
  states at $\sqrt{s} =$ 8 TeV}},  {\em JHEP} {\bf 08} (2014) 174,
  [\href{http://arxiv.org/abs/1405.3447}{{\tt arXiv:1405.3447}}].

\bibitem{Khachatryan:2016yji}
{\bf CMS} Collaboration, V.~Khachatryan et~al., {\it {Search for massive WH
  resonances decaying into the $\ell \nu\mathrm{ b \bar{b} }$ final state at
  $\sqrt{s}= $ 8 TeV}},  \href{http://arxiv.org/abs/1601.06431}{{\tt
  arXiv:1601.06431}}.

\bibitem{Khachatryan:2015bma}
{\bf CMS} Collaboration, V.~Khachatryan et~al., {\it {Search for a massive
  resonance decaying into a Higgs boson and a W or Z boson in hadronic final
  states in proton-proton collisions at $ \sqrt{s}=8 $ TeV}},  {\em JHEP} {\bf
  02} (2016) 145, [\href{http://arxiv.org/abs/1506.01443}{{\tt
  arXiv:1506.01443}}].

\bibitem{Khachatryan:2015ywa}
{\bf CMS} Collaboration, V.~Khachatryan et~al., {\it {Search for Narrow
  High-Mass Resonances in ProtonÐProton Collisions at $\sqrt{s}$ = 8 TeV
  Decaying to a Z and a Higgs Boson}},  {\em Phys. Lett.} {\bf B748} (2015)
  255--277, [\href{http://arxiv.org/abs/1502.04994}{{\tt arXiv:1502.04994}}].

\bibitem{Khachatryan:2015sja}
{\bf CMS} Collaboration, V.~Khachatryan et~al., {\it {Search for resonances and
  quantum black holes using dijet mass spectra in proton-proton collisions at
  $\sqrt{s} =$ 8 TeV}},  {\em Phys. Rev.} {\bf D91} (2015), no.~5 052009,
  [\href{http://arxiv.org/abs/1501.04198}{{\tt arXiv:1501.04198}}].

\bibitem{Dias:2015mhm}
F.~Dias, S.~Gadatsch, M.~Gouzevich, C.~Leonidopoulos, S.~Novaes, A.~Oliveira,
  M.~Pierini, and T.~Tomei, {\it {Combination of Run-1 Exotic Searches in
  Diboson Final States at the LHC}},
  \href{http://arxiv.org/abs/1512.03371}{{\tt arXiv:1512.03371}}.

\bibitem{Brehmer:2015dan}
J.~Brehmer et~al., {\it {The Diboson Excess: Experimental Situation and
  Classification of Explanations; A Les Houches Pre-Proceeding}},
  \href{http://arxiv.org/abs/1512.04357}{{\tt arXiv:1512.04357}}.

\bibitem{Hisano:2015gna}
J.~Hisano, N.~Nagata, and Y.~Omura, {\it {Interpretations of the ATLAS Diboson
  Resonances}},  {\em Phys. Rev.} {\bf D92} (2015), no.~5 055001,
  [\href{http://arxiv.org/abs/1506.03931}{{\tt arXiv:1506.03931}}].

\bibitem{Franzosi:2015zra}
D.~B. Franzosi, M.~T. Frandsen, and F.~Sannino, {\it {Diboson Signals via Fermi
  Scale Spin-One States}},  {\em Phys. Rev.} {\bf D92} (2015) 115005,
  [\href{http://arxiv.org/abs/1506.04392}{{\tt arXiv:1506.04392}}].

\bibitem{Cheung:2015nha}
K.~Cheung, W.-Y. Keung, P.-Y. Tseng, and T.-C. Yuan, {\it {Interpretations of
  the ATLAS Diboson Anomaly}},  {\em Phys. Lett.} {\bf B751} (2015) 188--194,
  [\href{http://arxiv.org/abs/1506.06064}{{\tt arXiv:1506.06064}}].

\bibitem{Dobrescu:2015qna}
B.~A. Dobrescu and Z.~Liu, {\it {W? Boson near 2 TeV: Predictions for Run 2 of
  the LHC}},  {\em Phys. Rev. Lett.} {\bf 115} (2015), no.~21 211802,
  [\href{http://arxiv.org/abs/1506.06736}{{\tt arXiv:1506.06736}}].

\bibitem{Gao:2015irw}
Y.~Gao, T.~Ghosh, K.~Sinha, and J.-H. Yu, {\it {SU(2)?SU(2)?U(1)
  interpretations of the diboson and Wh excesses}},  {\em Phys. Rev.} {\bf D92}
  (2015), no.~5 055030, [\href{http://arxiv.org/abs/1506.07511}{{\tt
  arXiv:1506.07511}}].

\bibitem{Thamm:2015csa}
A.~Thamm, R.~Torre, and A.~Wulzer, {\it {Composite Heavy Vector Triplet in the
  ATLAS Diboson Excess}},  {\em Phys. Rev. Lett.} {\bf 115} (2015), no.~22
  221802, [\href{http://arxiv.org/abs/1506.08688}{{\tt arXiv:1506.08688}}].

\bibitem{Brehmer:2015cia}
J.~Brehmer, J.~Hewett, J.~Kopp, T.~Rizzo, and J.~Tattersall, {\it {Symmetry
  Restored in Dibosons at the LHC?}},  {\em JHEP} {\bf 10} (2015) 182,
  [\href{http://arxiv.org/abs/1507.00013}{{\tt arXiv:1507.00013}}].

\bibitem{Cao:2015lia}
Q.-H. Cao, B.~Yan, and D.-M. Zhang, {\it {Simple non-Abelian extensions of the
  standard model gauge group and the diboson excesses at the LHC}},  {\em Phys.
  Rev.} {\bf D92} (2015), no.~9 095025,
  [\href{http://arxiv.org/abs/1507.00268}{{\tt arXiv:1507.00268}}].

\bibitem{Abe:2015uaa}
T.~Abe, T.~Kitahara, and M.~M. Nojiri, {\it {Prospects for Spin-1 Resonance
  Search at 13 TeV LHC and the ATLAS Diboson Excess}},  {\em JHEP} {\bf 02}
  (2016) 084, [\href{http://arxiv.org/abs/1507.01681}{{\tt arXiv:1507.01681}}].

\bibitem{Carmona:2015xaa}
A.~Carmona, A.~Delgado, M.~Quir—s, and J.~Santiago, {\it {Diboson resonant
  production in non-custodial composite Higgs models}},  {\em JHEP} {\bf 09}
  (2015) 186, [\href{http://arxiv.org/abs/1507.01914}{{\tt arXiv:1507.01914}}].

\bibitem{Allanach:2015hba}
B.~C. Allanach, B.~Gripaios, and D.~Sutherland, {\it {Anatomy of the ATLAS
  diboson anomaly}},  {\em Phys. Rev.} {\bf D92} (2015), no.~5 055003,
  [\href{http://arxiv.org/abs/1507.01638}{{\tt arXiv:1507.01638}}].

\bibitem{Dobrescu:2015yba}
B.~A. Dobrescu and Z.~Liu, {\it {Heavy Higgs bosons and the 2 TeV W$^{?}$
  boson}},  {\em JHEP} {\bf 10} (2015) 118,
  [\href{http://arxiv.org/abs/1507.01923}{{\tt arXiv:1507.01923}}].

\bibitem{Pelaggi:2015kna}
G.~M. Pelaggi, A.~Strumia, and S.~Vignali, {\it {Totally asymptotically free
  trinification}},  {\em JHEP} {\bf 08} (2015) 130,
  [\href{http://arxiv.org/abs/1507.06848}{{\tt arXiv:1507.06848}}].

\bibitem{Lane:2015fza}
K.~Lane and L.~Pritchett, {\it {Heavy Vector Partners of the Light Composite
  Higgs}},  {\em Phys. Lett.} {\bf B753} (2016) 211--214,
  [\href{http://arxiv.org/abs/1507.07102}{{\tt arXiv:1507.07102}}].

\bibitem{Faraggi:2015iaa}
A.~E. Faraggi and M.~Guzzi, {\it {Extra $Z^{\prime }$ s and $W^{\prime }$ s in
  heterotic-string derived models}},  {\em Eur. Phys. J.} {\bf C75} (2015),
  no.~11 537, [\href{http://arxiv.org/abs/1507.07406}{{\tt arXiv:1507.07406}}].

\bibitem{Low:2015uha}
M.~Low, A.~Tesi, and L.-T. Wang, {\it {Composite spin-1 resonances at the
  LHC}},  {\em Phys. Rev.} {\bf D92} (2015), no.~8 085019,
  [\href{http://arxiv.org/abs/1507.07557}{{\tt arXiv:1507.07557}}].

\bibitem{Dobrescu:2015jvn}
B.~A. Dobrescu and P.~J. Fox, {\it {Signals of a 2 TeV $W'$ boson and a heavier
  $Z'$ boson}},  \href{http://arxiv.org/abs/1511.02148}{{\tt
  arXiv:1511.02148}}.

\bibitem{Sajjad:2015urz}
A.~Sajjad, {\it {Understanding diboson anomalies}},  {\em Phys. Rev.} {\bf D93}
  (2016), no.~5 055028, [\href{http://arxiv.org/abs/1511.02244}{{\tt
  arXiv:1511.02244}}].

\bibitem{Das:2015ysz}
K.~Das, T.~Li, S.~Nandi, and S.~K. Rai, {\it {Diboson excesses in an anomaly
  free leptophobic left-right model}},  {\em Phys. Rev.} {\bf D93} (2016),
  no.~1 016006, [\href{http://arxiv.org/abs/1512.00190}{{\tt
  arXiv:1512.00190}}].

\bibitem{Deppisch:2015cua}
F.~F. Deppisch, L.~Graf, S.~Kulkarni, S.~Patra, W.~Rodejohann, N.~Sahu, and
  U.~Sarkar, {\it {Reconciling the 2 TeV excesses at the LHC in a linear seesaw
  left-right model}},  {\em Phys. Rev.} {\bf D93} (2016), no.~1 013011,
  [\href{http://arxiv.org/abs/1508.05940}{{\tt arXiv:1508.05940}}].

\bibitem{Dev:2015pga}
P.~S. Bhupal~Dev and R.~N. Mohapatra, {\it {Unified explanation of the $eejj$,
  diboson and dijet resonances at the LHC}},  {\em Phys. Rev. Lett.} {\bf 115}
  (2015), no.~18 181803, [\href{http://arxiv.org/abs/1508.02277}{{\tt
  arXiv:1508.02277}}].

\bibitem{Collins:2015wua}
J.~H. Collins and W.~H. Ng, {\it {A 2TeV W$_{R}$ , supersymmetry, and the Higgs
  mass}},  {\em JHEP} {\bf 01} (2016) 159,
  [\href{http://arxiv.org/abs/1510.08083}{{\tt arXiv:1510.08083}}].

\bibitem{ATLAS-CONF-2015-075}
{\it {Search for $WW/WZ$ resonance production in the $\ell\nu qq$ final state
  at $\sqrt{s}=13\,$ TeV with the ATLAS detector at the LHC}},  Tech. Rep.
  ATLAS-CONF-2015-075, CERN, Geneva, Dec, 2015.

\bibitem{ATLAS-CONF-2015-071}
{\it {Search for diboson resonances in the llqq final state in pp collisions at
  $\sqrt{s}$ = 13 TeV with the ATLAS detector}},  Tech. Rep.
  ATLAS-CONF-2015-071, CERN, Geneva, Dec, 2015.

\bibitem{ATLAS-CONF-2015-073}
{\it {Search for resonances with boson-tagged jets in 3.2 fb$^{-1}$ of $pp$
  collisions at $\sqrt{s}=13\,$ TeV collected with the ATLAS detector}},  Tech.
  Rep. ATLAS-CONF-2015-073, CERN, Geneva, Dec, 2015.

\bibitem{ATLAS-CONF-2015-068}
{\it {Search for diboson resonances in the $\nu\nu qq$ final state in $pp$
  collisions at $\sqrt{s}=$13 TeV with the ATLAS detector}},  Tech. Rep.
  ATLAS-CONF-2015-068, CERN, Geneva, Dec, 2015.

\bibitem{CMS-PAS-EXO-15-002}
{\bf CMS Collaboration} Collaboration, {\it {Search for massive resonances
  decaying into pairs of boosted W and Z bosons at $\sqrt{s}$ = 13 TeV}},
  Tech. Rep. CMS-PAS-EXO-15-002, CERN, Geneva, 2015.

\bibitem{ATLAS:2015nsi}
{\bf ATLAS} Collaboration, G.~Aad et~al., {\it {Search for new phenomena in
  dijet mass and angular distributions from $pp$ collisions at $\sqrt{s}=$ 13
  TeV with the ATLAS detector}},  {\em Phys. Lett.} {\bf B754} (2016) 302--322,
  [\href{http://arxiv.org/abs/1512.01530}{{\tt arXiv:1512.01530}}].

\bibitem{Coloma:2015una}
P.~Coloma, B.~A. Dobrescu, and J.~Lopez-Pavon, {\it {Right-handed neutrinos and
  the 2 TeV $W'$ boson}},  {\em Phys. Rev.} {\bf D92} (2015), no.~11 115023,
  [\href{http://arxiv.org/abs/1508.04129}{{\tt arXiv:1508.04129}}].

\bibitem{Martin:2009iq}
A.~D. Martin, W.~J. Stirling, R.~S. Thorne, and G.~Watt, {\it {Parton
  distributions for the LHC}},  {\em Eur. Phys. J.} {\bf C63} (2009) 189--285,
  [\href{http://arxiv.org/abs/0901.0002}{{\tt arXiv:0901.0002}}].

\bibitem{Cao:2012ng}
Q.-H. Cao, Z.~Li, J.-H. Yu, and C.~P. Yuan, {\it {Discovery and Identification
  of W' and Z' in SU(2) x SU(2) x U(1) Models at the LHC}},  {\em Phys. Rev.}
  {\bf D86} (2012) 095010, [\href{http://arxiv.org/abs/1205.3769}{{\tt
  arXiv:1205.3769}}].

\bibitem{Carena:2004xs}
M.~Carena, A.~Daleo, B.~A. Dobrescu, and T.~M.~P. Tait, {\it {$Z^\prime$ gauge
  bosons at the Tevatron}},  {\em Phys. Rev.} {\bf D70} (2004) 093009,
  [\href{http://arxiv.org/abs/hep-ph/0408098}{{\tt hep-ph/0408098}}].

\bibitem{Hamberg:1990np}
R.~Hamberg, W.~L. van Neerven, and T.~Matsuura, {\it {A Complete calculation of
  the order $\alpha-s^{2}$ correction to the Drell-Yan $K$ factor}},  {\em
  Nucl. Phys.} {\bf B359} (1991) 343--405. [Erratum: Nucl.
  Phys.B644,403(2002)].

\bibitem{Agashe:2014kda}
{\bf Particle Data Group} Collaboration, K.~A. Olive et~al., {\it {Review of
  Particle Physics}},  {\em Chin. Phys.} {\bf C38} (2014) 090001.

\bibitem{Bellomo_Moriond}
M.~Bellomo, {\it {Moriond 2016: Searches for Boosted Di-Boson Resonances with
  the ATLAS and CMS detectors}}, .

\bibitem{ATLAS-CONF-2015-074}
{\it {Search for new resonances decaying to a W or Z boson and a Higgs boson in
  the $\ell\ell b\bar b$, $\ell\nu b\bar b$, and $\nu\nu b\bar b$ channels in
  $pp$ collisions at $\sqrt s = 13$~TeV with the ATLAS detector}},  Tech. Rep.
  ATLAS-CONF-2015-074, CERN, Geneva, Dec, 2015.

\bibitem{Aad:2014cka}
{\bf ATLAS} Collaboration, G.~Aad et~al., {\it {Search for high-mass dilepton
  resonances in pp collisions at $\sqrt{s}=8$  TeV with the ATLAS
  detector}},  {\em Phys. Rev.} {\bf D90} (2014), no.~5 052005,
  [\href{http://arxiv.org/abs/1405.4123}{{\tt arXiv:1405.4123}}].

\bibitem{Khachatryan:2014fba}
{\bf CMS} Collaboration, V.~Khachatryan et~al., {\it {Search for physics beyond
  the standard model in dilepton mass spectra in proton-proton collisions at $
  \sqrt{s}=8 $ TeV}},  {\em JHEP} {\bf 04} (2015) 025,
  [\href{http://arxiv.org/abs/1412.6302}{{\tt arXiv:1412.6302}}].

\bibitem{Khachatryan:2015dcf}
{\bf CMS} Collaboration, V.~Khachatryan et~al., {\it {Search for narrow
  resonances decaying to dijets in proton-proton collisions at $\sqrt(s) =$ 13
  TeV}},  {\em Phys. Rev. Lett.} {\bf 116} (2016), no.~7 071801,
  [\href{http://arxiv.org/abs/1512.01224}{{\tt arXiv:1512.01224}}].

\bibitem{Han:2004az}
Z.~Han and W.~Skiba, {\it {Effective theory analysis of precision electroweak
  data}},  {\em Phys. Rev.} {\bf D71} (2005) 075009,
  [\href{http://arxiv.org/abs/hep-ph/0412166}{{\tt hep-ph/0412166}}].

\bibitem{Han:2005pr}
Z.~Han, {\it {Electroweak constraints on effective theories with U(2) x (1)
  flavor symmetry}},  {\em Phys. Rev.} {\bf D73} (2006) 015005,
  [\href{http://arxiv.org/abs/hep-ph/0510125}{{\tt hep-ph/0510125}}].

\bibitem{Efrati:2015eaa}
A.~Efrati, A.~Falkowski, and Y.~Soreq, {\it {Electroweak constraints on
  flavorful effective theories}},  {\em JHEP} {\bf 07} (2015) 018,
  [\href{http://arxiv.org/abs/1503.07872}{{\tt arXiv:1503.07872}}].

\bibitem{CMS-PAS-EXO-15-005}
{\bf CMS Collaboration} Collaboration, {\it {Search for a Narrow Resonance
  Produced in 13 TeV pp Collisions Decaying to Electron Pair or Muon Pair Final
  States}},  Tech. Rep. CMS-PAS-EXO-15-005, CERN, Geneva, 2015.

\end{thebibliography}\endgroup


\end{document}